%% file: ANA-FTAG-2019-02-PAPER.tex
\pdfoutput=1
\documentclass[cernpreprint, texlive=2023, UKenglish, texmf, orcidlogo]{atlasdoc}
 
\usepackage[subcaption]{atlaspackage}
\usepackage{tikz}
\definecolor{mydarkgreen}{rgb}{0.30,0.5,0.18}
\usepackage{atlasbiblatex}
 
\usepackage{atlasphysics}

\graphicspath{{logos/}{figures/}}

\AtlasTitle{Calibration of the light-flavour jet mistagging efficiency of the $b$-tagging algorithms with $Z$+jets events using 139~$\mathrm{fb}^{-1}$ of ATLAS proton--proton collision data at $\sqrt{s} = 13$~TeV}

\AtlasAbstract{
The identification of $b$-jets, referred to as $b$-tagging, is an important part of many physics analyses in the ATLAS experiment at the Large Hadron Collider and an accurate calibration of its performance is essential for high-quality physics results.
This publication describes the calibration of the light-flavour jet mistagging efficiency in a data sample of proton--proton collision events at $\sqrt{s}=13$~TeV corresponding to an integrated luminosity of 139~fb$^{-1}$. The calibration is performed in a sample of $Z$ bosons produced in association with jets.
Due to the low mistagging efficiency for light-flavour jets, a method which uses modified versions of the $b$-tagging algorithms referred to as flip taggers is used in this work. A fit to the jet-flavour-sensitive secondary-vertex mass is performed to extract a scale factor from data, to correct the light-flavour jet mistagging efficiency in Monte Carlo simulations, while simultaneously correcting the $b$-jet efficiency. With this procedure, uncertainties coming from the modeling of jets from heavy-flavour hadrons are considerably lower than in previous calibrations of the mistagging scale factors, where they were dominant.
The scale factors obtained in this calibration are consistent with unity within uncertainties.
}

\AtlasRefCode{FTAG-2019-02}
 
\PreprintIdNumber{CERN-EP-2022-211}

\AtlasJournal{EPJC}
\AtlasJournalRef{\EPJC 83 (2023) 728}
\AtlasDOI{10.1140/epjc/s10052-023-11736-z}

\hypersetup{pdftitle={ATLAS document},pdfauthor={The ATLAS Collaboration}}
 
\begin{document}
 
\maketitle

\section{Introduction}

Many analyses in ATLAS \cite{PERF-2007-01}, such as measurements or searches involving top quarks or Higgs bosons, rely on the identification of jets containing $b$-hadrons ($b$-jets) with high tagging efficiency and low mistagging efficiency for jets containing $c$-hadrons ($c$-jets) or containing neither $b$- nor $c$-hadrons (light-flavour jets).
The relatively long lifetime and high mass of $b$-hadrons together with the large track multiplicity of their decay products is exploited by $b$-tagging algorithms to identify $b$-jets.
 
The $b$-tagging algorithms are trained using Monte Carlo (MC) simulated events and therefore need to be calibrated in order to correct for efficiency differences between data and simulation that may arise from an imperfect description of the data, e.g. in the parton shower and fragmentation modelling or in the detector and response simulation.
The efficiency of identifying a $b$-jet ($\varepsilon^{b}$) and the mistagging efficiencies ($\varepsilon^{c}$ and $\varepsilon^{\text{light}}$), which are the probabilities that other jets are wrongly identified by the algorithms as $b$-jets, are measured in data and compared with the predictions of the simulation.
These tagging and mistagging efficiencies are defined as
\begin{equation*}
\varepsilon^{f} = \frac{N_{\mathrm{pass}}^{f}}{N_{\mathrm{all}}^{f}},
\end{equation*}
where $f \in \{b,~c,~\text{light}\}$, $N_{\mathrm{pass}}^{f}$ is the number of jets of flavour $f$ selected by the $b$-tagging algorithm and $N_{\mathrm{all}}^{f}$ is the number of all jets of flavour $f$.
The flavour-tagging efficiency in data ($\varepsilon^{f}_{\mathrm{data}}$) is compared with the efficiency in MC simulation ($\varepsilon^{f}_{\mathrm{MC}}$) and a calibration factor, also called the scale factor (SF), is defined as
\begin{equation}\label{eq:SF}
\mathrm{SF}^{f} = \frac{\varepsilon^{f}_{\mathrm{data}}}{\varepsilon^{f}_{\mathrm{MC}}}.
\end{equation}
The calibration factors correct the efficiencies and mistagging efficiencies in simulation to the ones in data and are applied to all physics analyses in ATLAS that use $b$-tagging.
The $b$-tagging efficiencies and the calibration SFs depend on the jet kinematics.
The $b$-jet efficiency ($\varepsilon^{b}$) is calibrated using the method described in Ref.~\cite{FTAG-2018-01}, where the SFs$^{b}$ are extracted from a sample of events containing top-quark pairs decaying into a final state with two charged leptons and two $b$-jets. 
The $c$-jet mistagging efficiency ($\varepsilon^{c}$) is calibrated via the method described in Ref.~\cite{FTAG-2020-08}, where the SFs$^{c}$ are extracted from events containing top-quark pairs decaying into a final state with exactly one charged lepton and several jets. The events are reconstructed using a kinematic likelihood technique and include a hadronically decaying $W$ boson, whose decay products are rich in $c$-jets.
 
This paper describes the measurement of the light-flavour jet mistagging efficiency ($\varepsilon^{\text{light}}$) of the \texttt{DL1r} $b$-tagger, which is widely used in ATLAS Run~2 physics analyses and is discussed in Section~\ref{sec:btag}. The SFs$^{\text{light}}$ are extracted from particle-flow jets~\cite{PERF-2015-09} using 139~$\mathrm{fb}^{-1}$ of proton--proton (\pp) collision data collected by the ATLAS detector during Run~2 of the LHC.
 
The mistagging efficiency $\varepsilon^{\text{light}}$ is difficult to calibrate because after applying a $b$-tagging requirement, the obtained sample of jets is strongly dominated by $b$-jets and the fraction of light-flavour jets passing a selection on the $b$-tagging score is too low to estimate $\varepsilon^{\text{light}}_{\text{data}}$.
In order to extract an unbiased and precise SF$^{\text{light}}$, a sample enriched in mistagged light-flavour jets is required.
 
In this paper, the Negative Tag method~\cite{ATLAS-CONF-2014-046,D0:2006hcm}, which relies on a modified tagger with a reduced $\varepsilon^{b}$ and $\varepsilon^{c}$ but similar $\varepsilon^{\text{light}}$ with respect to the nominal tagger, is used. The method, described in detail in Section~\ref{sec:method}, has already been used to calibrate $\varepsilon^{\text{light}}$ by using dijet events in 2015--2016 data~\cite{ATLAS-CONF-2018-006} from the early part of Run~2.
 
In this work, jets produced in association with $Z$ bosons ($Z$+jets) are used instead, allowing the use of unprescaled lepton triggers instead of the prescaled single-jet triggers which were used in the previous calibration.
The calibration precision is improved by extracting the SF$^{\text{light}}$ in a fit that simultaneously corrects $\varepsilon^{b}$ using data. Previously, $\varepsilon^{b}$ was estimated from simulation, resulting in large modelling uncertainties which impacted the precision of the result.
An alternative approach~\cite{ATLAS-CONF-2018-006} to the Negative Tag method exists in which $\varepsilon^{\text{light}}_{\text{data}}$ is estimated from simulation by applying data-driven corrections to the input quantities of the low-level $b$-tagging algorithms. This method is used to assess the extrapolation uncertainty between the modified and nominal tagger as described in Section \ref{sec:unc}.
 
The paper is structured as follows. Section~\ref{sec:detector} describes the ATLAS detector. Section~\ref{sec:samples} presents the dataset and simulations used in this calibration. The reconstruction of jets, electrons and muons is summarised in Section~\ref{sec:selection}, while the $b$-tagging algorithms are described in detail in Section~\ref{sec:btag}. The calibration method is detailed in Section~\ref{sec:method}. Systematic uncertainties and results are presented, respectively, in Section~\ref{sec:unc} and Section~\ref{sec:result} and conclusions are given in Section~\ref{sec:conclusions}.


\FloatBarrier

\section{ATLAS detector}
\label{sec:detector}

\newcommand{\AtlasCoordFootnote}{
ATLAS uses a right-handed coordinate system with its origin at the nominal interaction point in the centre of the detector and the \(z\)-axis along the beam pipe. The \(x\)-axis points from the interaction point to the centre of the LHC ring, and the \(y\)-axis points upwards. Cylindrical coordinates \((r,\phi)\) are used in the transverse plane, \(\phi\) being the azimuthal angle around the \(z\)-axis. The pseudorapidity is defined in terms of the polar angle \(\theta\) as \(\eta = -\ln \tan(\theta/2)\). The angular distance is measured in units of \(\Delta R \equiv \sqrt{(\Delta\eta)^{2} + (\Delta\phi)^{2}}\).}
 
The ATLAS experiment~\cite{PERF-2007-01} at the LHC is a multipurpose particle detector with a forward--backward symmetric cylindrical geometry and a near \(4\pi\) coverage in solid angle.\footnote{\AtlasCoordFootnote} It consists of an inner tracking detector (ID) surrounded by a thin superconducting solenoid providing a \SI{2}{\tesla} axial magnetic field, electromagnetic and hadron calorimeters, and a muon spectrometer with a toroidal magnet system.
 
The inner tracking detector, which provides full coverage of a pseudorapidity range \(|\eta| < 2.5\), consists of silicon pixel, silicon microstrip, and transition radiation tracking detectors. The high-granularity silicon pixel detector covers the vertex region and typically provides four measurements per track, the first measurement normally being in the insertable B-layer (IBL) installed before Run~2, primarily to enhance the $b$-tagging performance~\cite{ATLAS-TDR-19,PIX-2018-001}.
Sampling calorimeters, made of lead and liquid Argon (LAr), provide electromagnetic (EM) energy measurements with high granularity in the pseudorapidity region $|\eta| < 3.2$.
A steel/scintillator-tile hadron calorimeter covers the central pseudorapidity range (\(|\eta| < 1.7\)) and a copper/LAr hadron calorimeter covers the range $1.5 < |\eta| < 3.2$.
The forward region is instrumented with LAr calorimeters in the range $3.1 < |\eta| < 4.9$, measuring both electromagnetic and hadronic energies in copper/LAr and tungsten/LAr modules.
The muon spectrometer surrounds the calorimeters and is based on three large superconducting air-core toroidal magnets with eight coils each. The field integral of the toroids ranges between \num{2.0} and \SI{6.0}{\tesla\metre} across most of the detector. The muon spectrometer includes a system of precision tracking chambers and fast detectors for triggering.
 
A two-level trigger system is used to select events. The first-level trigger is implemented in hardware and uses a subset of the detector information to accept events at a rate below \SI{100}{\kHz}. This is followed by a software-based trigger that reduces the accepted event rate to \SI{1}{\kHz} on average depending on the data-taking conditions~\cite{ATLAS:2016wtr}.
 
An extensive software suite~\cite{ATL-SOFT-PUB-2021-001} is used in data simulation, in the reconstruction and analysis of real and simulated data, in detector operations, and in the trigger and data acquisition systems of the experiment.


\FloatBarrier

\section{Data and simulated samples}
\label{sec:samples}

The presented results are based on data from \pp collisions with a centre-of-mass energy of $\sqrt{s}=13$~\TeV\ and a 25~ns proton bunch crossing interval. The data were collected between 2015 and 2018 with the ATLAS detector, corresponding to an integrated luminosity of 139~\ifb~\cite{DAPR-2021-01,LUCID2}. A set of single-electron~\cite{TRIG-2018-05} and single-muon triggers~\cite{TRIG-2018-01} were used, with transverse momentum (\pt) thresholds in the range 20--26~\GeV\ and depending on the lepton flavour and data-taking period. All detector subsystems are required to have been operational during data taking and to fulfil data quality requirements~\cite{DAPR-2018-01}.
 
Expected Standard Model (SM) processes were modelled using MC simulation. A summary of the simulated samples used in this calibration and their basic parameter settings is provided in Table~\ref{tab:mc}. All samples were produced using the ATLAS simulation infrastructure~\cite{SOFT-2010-01} and \GEANT~\cite{Agostinelli:2002hh}. The effect of additional \pp interactions per bunch crossing (pile-up) is accounted for by overlaying the hard-scattering process with minimum-bias events generated with \Pythia[8.186]~\cite{Sjostrand:2007gs} using the \NNPDF[2.3lo] set of parton distribution functions (PDF)~\cite{Ball:2012cx} and the A3 set of tuned parameters~\cite{ATL-PHYS-PUB-2016-017}. Different pile-up conditions between data and simulation are taken into account by reweighting the mean number of interactions per bunch crossing in simulation to the number observed in data.
 
The events used in this study originate mostly from the production of a \Zboson boson in association with jets. This process is simulated with two different generators.
Nominal predictions of $Z/\gamma^{*}$ production were generated using \Sherpa[2.2.1]~\cite{Gleisberg:2008ta} with the \NNPDF[3.0nnlo] PDF~\cite{Ball:2014uwa}. Diagrams with up to two additional parton emissions were simulated with next-to-leading-order (\NLO) precision in QCD, and those with three or four additional parton emissions to leading-order accuracy. Matrix elements were merged with the \Sherpa parton shower using the MEPS@NLO formalism~\cite{Catani:2001cc,Hoeche:2012yf,Hoeche:2009rj,Hoeche:2011fd} with a merging scale of 20~\GeV\ and the five-flavour numbering scheme (5FNS).
The sample used to evaluate the tagging efficiencies predicted by the MC simulation was produced with the \MADGRAPH[2.2.3]~\cite{Alwall:2014hca} event generator with up to four additional partons at matrix-element level interfaced with \Pythia[8.210] to model the parton shower, hadronisation, and underlying event with the \NNPDF[3.0nlo] PDF for the processes where the \Zboson boson decays into muons or electrons, and with \MADGRAPH[2.2.2] interfaced with \Pythia[8.186] with the \NNPDF[2.3lo] PDF~\cite{Ball:2012cx} for \Zboson boson decays into $\tau$-leptons. The predictions of the $Z$+jets \MADGRAPH program are based on calculations using the 5FNS scheme.
\Sherpa~2.2.1 and \MADGRAPH $Z$+jets simulations are normalised to the overall cross-section prediction at NNLO precision in QCD~\cite{Anastasiou:2003ds}.

In addition to $Z$+jets production, some minor backgrounds contribute to the final event sample used for the calibration. These backgrounds consist mostly of diboson production ($\ensuremath{ZZ}\xspace$ and $\ensuremath{ZW}\xspace$) and top-quark pair production with subsequent decays into final states with two charged leptons.
The $\ensuremath{ZW}\xspace$ production and the $\ensuremath{ZZ}\xspace$ production via quark--antiquark annihilation were simulated with \Sherpa[2.2.1] with a MEPS@NLO configuration similar to what was used in the simulation of the $Z$+jets process described above, while loop-induced $ZZ$ processes initiated by gluon--gluon fusion were generated with \Sherpa[2.2.2].
Both the $\ensuremath{ZZ}\xspace$ and $\ensuremath{ZW}\xspace$ simulation samples used the \NNPDF[3.0nnlo] PDF set.
Top-quark pair production was simulated using the \POWHEGBOX[v2] generator at \NLO precision in QCD~\cite{Frixione:2007vw} interfaced with \Pythia[8.230] to model the parton shower, hadronisation, and underlying event, using the A14 set of tuned parameters~\cite{ATL-PHYS-PUB-2014-021}.
This simulation was used to model top-quark pair production with subsequent decays into final states with two charged leptons and into final states with one charged lepton.
The latter is used to evaluate the difference between the SFs obtained from the modified (\texttt{DL1rFlip}) and nominal (\texttt{DL1r}) $b$-tagging algorithms.
 
The decays of $b$- and $c$-hadrons were handled by \EVTGEN[1.6.0]~\cite{Lange:2001uf} in all simulations, except for those generated using \Sherpa, for which the default configuration recommended by the \Sherpa authors was used.
 
\begin{table}
\caption{A summary of MC generators and their basic parameter settings used to simulate the various physics processes.}
\label{tab:mc}
{
\begin{center}
\begin{tabular}{lccc}
\toprule
Process & Generator & PDF set & Hadronisation/Fragmentation \\
\midrule
$Z$+jets & \Sherpa~2.2.1 & \NNPDF[3.0] & \Sherpa~2.2.1  \\
$\ensuremath{ZW}\xspace$ and \qqbar-initiated $\ensuremath{ZZ}\xspace$ & \Sherpa~2.2.1 & \NNPDF[3.0] & \Sherpa~2.2.1 \\
$\ensuremath{ZZ}\xspace$ from gluon--gluon fusion & \Sherpa~2.2.2 & \NNPDF[3.0] & \Sherpa~2.2.2  \\
\ttbar & \POWHEGBOX~v2 & \NNPDF[3.0]  & \Pythia[8.230]  \\
$Z$+jets ($Z \rightarrow ee$ and $Z \rightarrow \mu\mu$) & \MADGRAPH 2.2.3 & \NNPDF[3.0] & \Pythia[8.210] \\
$Z$+jets ($Z \rightarrow \tau\tau$) & \MADGRAPH 2.2.2 & \NNPDF[2.3] & \Pythia[8.186] \\
\bottomrule
\end{tabular}
\end{center}
}
\end{table}


\FloatBarrier
 
\section{Object and event selection}
\label{sec:selection}

Tracks of charged particles are reconstructed from energy deposits in the material of the ID~\cite{PERF-2015-08}.
Events are required to contain at least one vertex with two or more associated tracks that must have $\pT > 500$~\MeV. 
Among all vertices, the vertex with the highest $p_{\text{T}}^2$ sum of the associated tracks is taken as the primary vertex (PV) \cite{ATL-PHYS-PUB-2015-026}.
The transverse ($d_{0}$) track impact parameter (IP) is defined as the distance of closest approach of the track-trajectory to the PV in the transverse plane. The longitudinal track IP ($z_{0}$) is defined as the distance in the $z$-direction between the PV and the track trajectory at the point of the closest approach in the $x$--$y$ plane.
 
Jets containing $b$- and $c$-hadrons are characterised by having in-flight decays due to the relatively long lifetime of the heavy-flavour hadrons and give rise to secondary vertices (SV) with associated tracks. To exploit the presence of in-flight decays in the jet direction, the signed IP is defined for each track within a jet. As shown in Figure \ref{fig:IP}, a track has a positive IP if the angle between the jet direction and the line joining the PV to the point of closest approach to the track is less than $\pi/2$ and negative otherwise. The Single Secondary Vertex Finder (\texttt{SSVF}) algorithm~\cite{ATL-PHYS-PUB-2017-011} is used for the reconstruction of the SV.
 
\begin{figure}[h!]
\center
\includegraphics[]{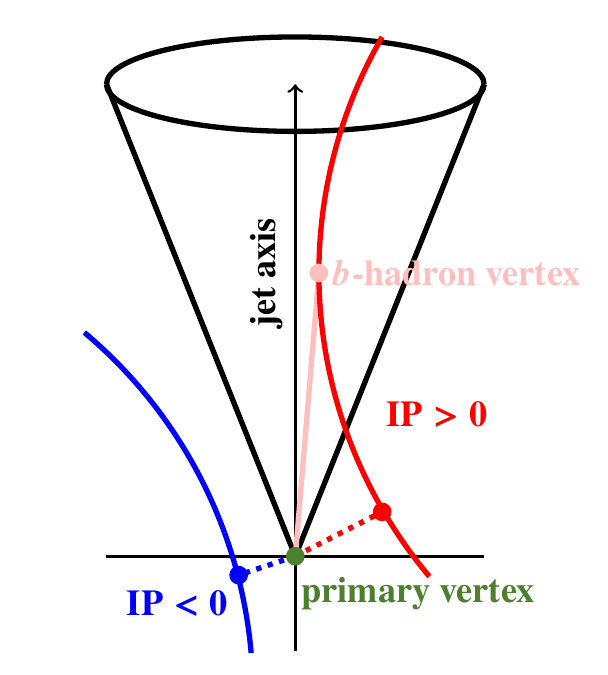}
\caption{Graphical illustration of the signed IP of tracks in the jet.}
\label{fig:IP}
\end{figure}
 
Electron candidates are reconstructed from energy clusters in the EM calorimeter matched to an ID track~\cite{EGAM-2018-01}.
Candidates must satisfy $\pT > 18$~GeV\ and $|\eta| < 2.47$, excluding the barrel--endcap transition region of 1.37$< |\eta| < $1.52, and their associated tracks must fulfil $|d_{0}|/\sigma_{d_{0}} < 5$ and $|z_{0}|\sin\theta < 0.5$~mm, where $\sigma_{d_{0}}$ is the uncertainty in $d_{0}$.
Electrons are identified using a likelihood-based discriminant that combines information about shower shapes in the EM calorimeter, track properties and the quality of the track-to-cluster matching~\cite{EGAM-2018-01}.
Electrons must fulfill a `Tight' identification selection.
The gradient isolation criteria are applied to reject non-prompt electrons, using energy deposits in the ID and calorimeter within a cone around the electron. Their efficiency is 90\% (99\%) for electrons from $Z\rightarrow ee$ decays at $\pT=25~(60)$~GeV.
Efficiency scale factors measured in data~\cite{EGAM-2018-01} are used to correct for differences in reconstruction, identification, isolation and trigger selection efficiencies between data and simulation.
 
Muon candidates are reconstructed by combining tracks in the ID with tracks in the muon spectrometer~\cite{PERF-2015-10}.
Only muons with $|\eta| < 2.5$ and $\pT > 18$~GeV\ are used.
They have to fulfil the `Tight' identification selection criteria~\cite{MUON-2018-03} and satisfy $|d_{0}| / \sigma_{d_{0}} < 3$ and $|z_{0}|\sin\theta <0.5$~mm.
The isolation requirement reduces the contamination from non-prompt muons by placing an upper bound on the amount of energy measured in the tracking detectors and the calorimeter (combined using the particle-flow algorithm~\cite{PERF-2015-09}) within a cone of variable size (for $\pT^\mu < 50$~\GeV) or fixed size (for $\pT^\mu > 50$~\GeV) around the muon.
Efficiency scale factors are used to correct for differences in muon reconstruction, identification, vertex association, isolation and trigger efficiencies between simulation and data~\cite{MUON-2018-03}.
 
Jets are reconstructed from particle-flow objects combining information from the tracker and calorimeter~\cite{PERF-2015-09}, using the anti-$k_{t}$ algorithm ~\cite{Cacciari:2008gp,Fastjet} with a radius parameter of $R=0.4$.
The jet energy is calibrated to the particle scale by using a sequence of corrections, including simulation-based corrections and in situ calibrations~\cite{JETM-2018-05}. 
The jet-vertex tagging technique (JVT)~\cite{PERF-2014-03}, which uses a multivariate likelihood approach, is applied to jets with $|\eta| < 2.4$ and $\pT < 60$~\GeV\ to suppress jets from pile-up activity. The `Tight' selection criterion, corresponding to a JVT score $>$ 0.5, is used. Scale factors are applied to match the JVT MC efficiencies to those in data.
All selected jets must have a $\pT > 20$~\GeV\ and $|\eta|< 2.5$.
 
The $b$-tagging algorithms use tracks matched to jets as input. This matching uses the angular separation between the track momenta, defined at the point of closest approach to the PV, and the jet axis, $\Delta R(\mathrm{track,jet~axis})$. The selection requirement on $\Delta R(\mathrm{track,jet~axis})$ varies as a function of the jet \pT because the $b$-hadron decay products are more collimated at larger hadron \pT~\cite{FTAG-2018-01}.
 
The jet-flavour labelling in simulation is based on an angular matching of reconstructed jets to generator-level $b$-hadrons, $c$-hadrons and $\tau$-leptons with $\pT > 5$~\GeV. If a $b$-hadron is found within a cone of size $\Delta R = 0.3$ around the jet axis, the jet is labelled as a $b$-jet. If no matching to any $b$-hadron is possible, the matching procedure is repeated sequentially for $c$-hadrons and $\tau$-leptons, and the matched jets are called $c$-jets and $\tau$-jets, respectively. A jet is labelled as a light-flavour jet by default if no matching to any of these particles was successful.
 
A series of requirements on the angular separation $\Delta R$ between muons, electrons and jets are applied to remove overlaps between objects. 
If an electron candidate shares an ID track with a muon candidate, the electron candidate is rejected.  Jets within a cone of $\Delta R = 0.2$ around a lepton are rejected, unless the lepton is a muon and the jet has more than three associated tracks, in which case the muon is rejected. Finally, lepton candidates that are found to be $0.2 < \Delta R < 0.4$ from any remaining jet are discarded.
 
A jet sample enriched in light-flavour jets is needed for the calibration.
A sample constructed using the leading jet in \pT in a $Z$+jets selection is expected to contain a 7\%--8\% fraction of $c$-jets and a 4\%--6\% fraction of $b$-jets, depending on the jet \pT.
The mistagging efficiency calibration is performed on the leading jet in a sample of $Z$+jets candidate events.
The distinct signature of a $Z$ boson decaying into either two electrons or two muons allows a clean sample of $Z$+jets events to be selected.
Events are selected for further analysis using single-lepton triggers, where one of the leptons must be matched to an object that triggered the recording of the event.
The event must contain exactly two prompt leptons with opposite charges and the same flavour (i.e. either exactly two electrons or exactly two muons). The leading lepton must have $\pT > 28$~\GeV\ and the invariant mass of the dilepton system, $m_{\ell\ell}$, must satisfy $81 < m_{\ell\ell} < 101$~\GeV.
Only events with a reconstructed $Z$ boson with $p_{\mathrm{T}}^{Z} > 50$~\GeV\ are considered because the overall modelling is better in this range~\cite{ATLAS:2018sgt}.
The event must also contain at least one jet with $\pT > 20$~\GeV\ and $|\eta| < 2.5$.
 
The event yields obtained after applying all selection requirements are listed in Table~\ref{tab:Yields}, while comparisons between the predictions by the simulation and the data are shown in Figure~\ref{fig:DataMC} for selected variables.
The predictions of the MC simulation are normalised to data for these comparisons.
The shapes of the distributions predicted by the simulation are consistent with those observed within statistical uncertainties for most of the considered kinematic range.
The discrepancy observed in the SV mass ($m_{\mathrm{SV}}$) distribution is expected because it is affected by a known mismodelling of the jet-flavour fractions~\cite{STDM-2017-38,CMS:2020cso}; this is mitigated by the fit described in Section \ref{sec:method}. The $m_{\mathrm{SV}}$ observable is calculated using the \texttt{SSVF} algorithm. Negative values of $m_{\mathrm{SV}}$ indicate that no SV was found in the jet. The mismodelling in jet \pT does not affect the calibration.
 
\begin{table}[!htbp] 
\caption{Expected contributions of processes in the calibration region, compared with the selected data obtained with the full Run~2 dataset. The quoted uncertainties are statistical uncertainties only. Normalisation differences between data and MC simulation are taken into account in the calibration strategy and do not affect the calibration results.}
\label{tab:Yields}

 
{
\begin{center}
\begin{tabular}{l c }
\toprule
Process & Yield \\
\midrule
$Z$+jets (\Sherpa) & (8.766 $\pm$ 0.015)$\cdot 10^{6}$ \\
$t\bar{t}$ &  (7.673 $\pm$  0.006)$\cdot 10^{4}$ \\
$WZ$ & (4.188 $\pm$ 0.005)$\cdot 10^{4}$  \\
$ZZ$ &  (2.526 $\pm$ 0.004)$\cdot 10^{4}$\\
\midrule
Predictions & (8.910 $\pm$ 0.015)$\cdot 10^{6}$  \\
Data &  9.560 $\cdot 10^{6}$   \\
\bottomrule
\end{tabular}
\end{center}
}


\end{table}

\begin{figure}[h!]
\centering
\includegraphics[width=0.4\textwidth]{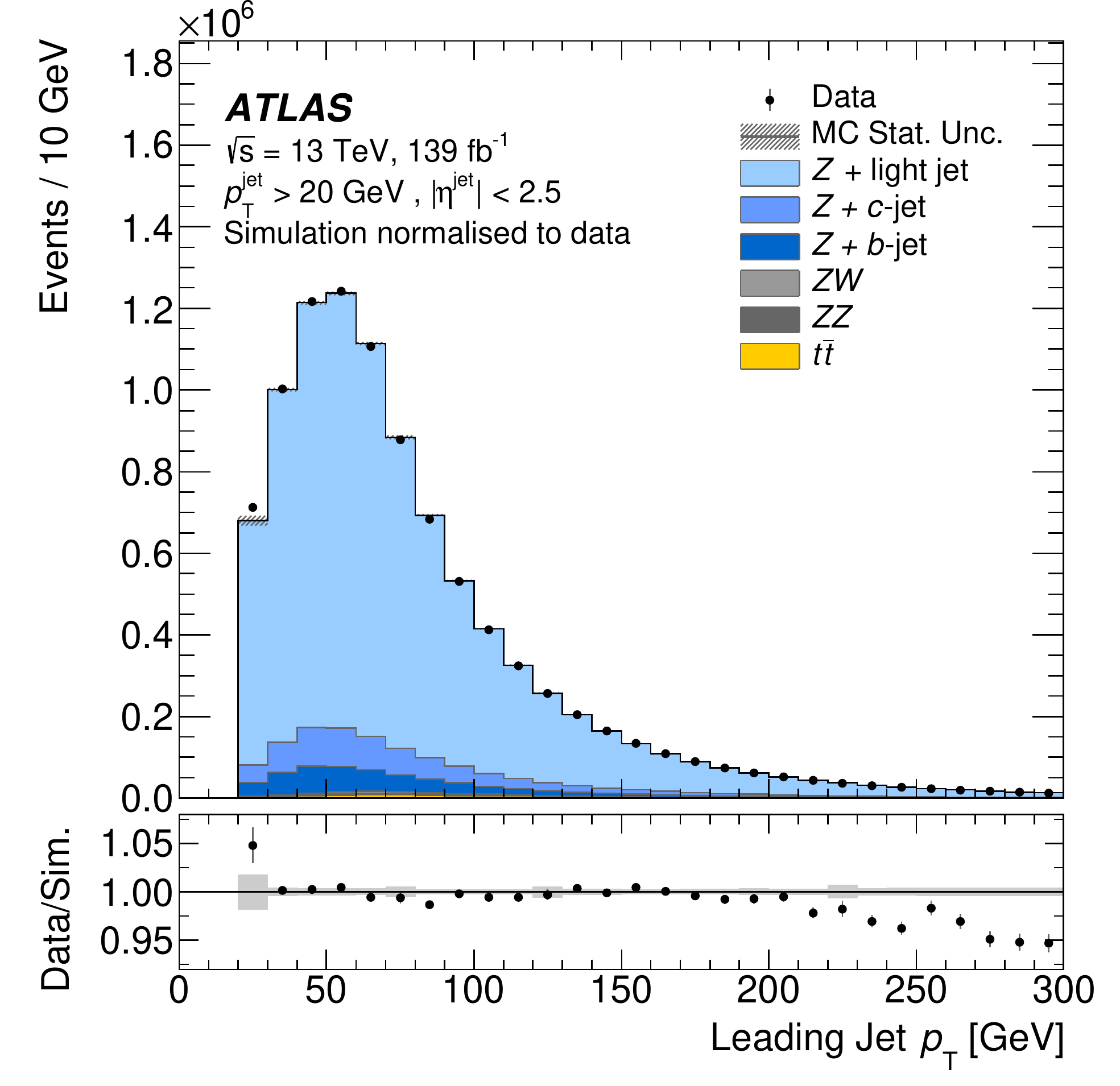}
\includegraphics[width=0.4\textwidth]{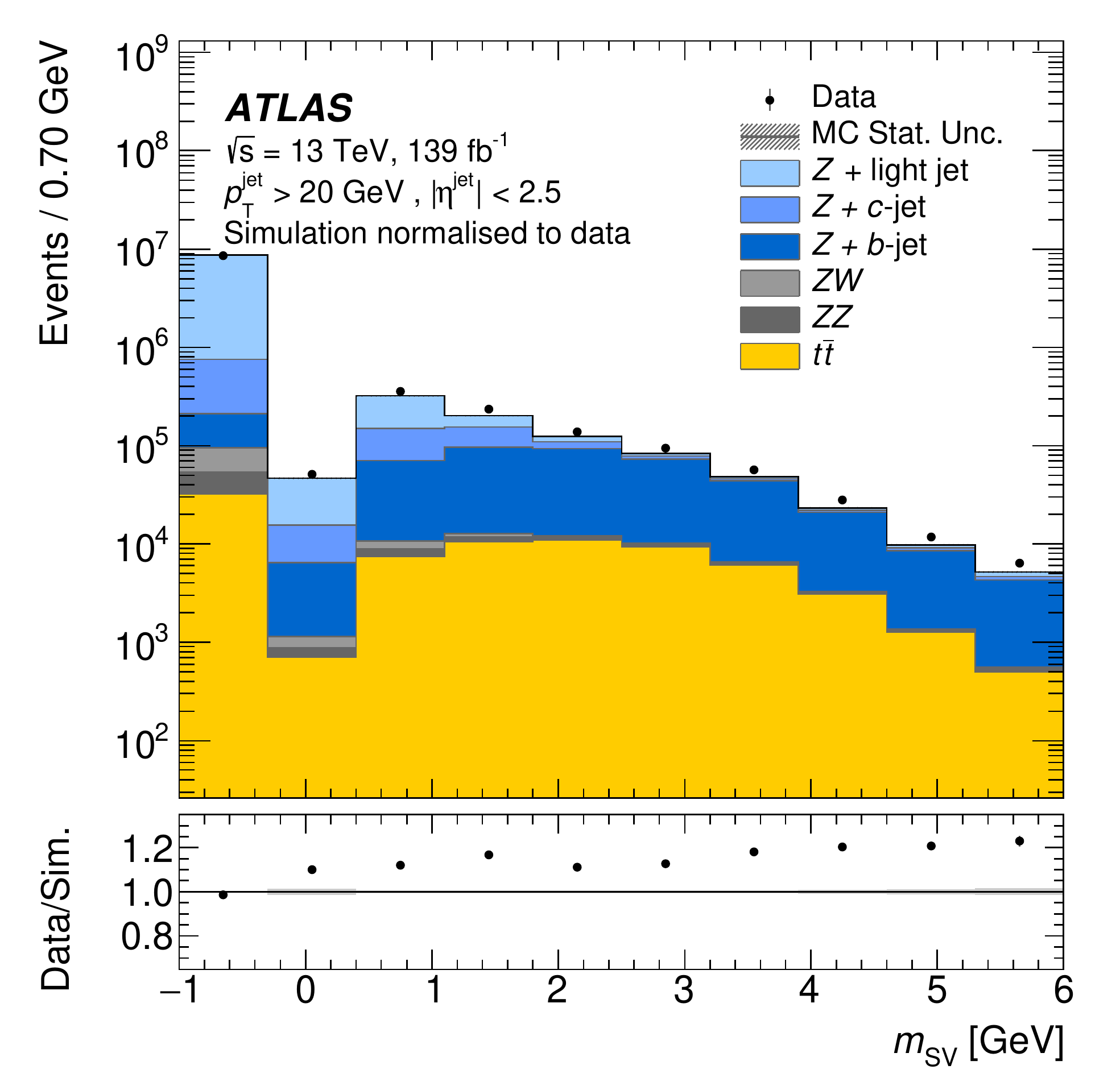}\\
\includegraphics[width=0.4\textwidth]{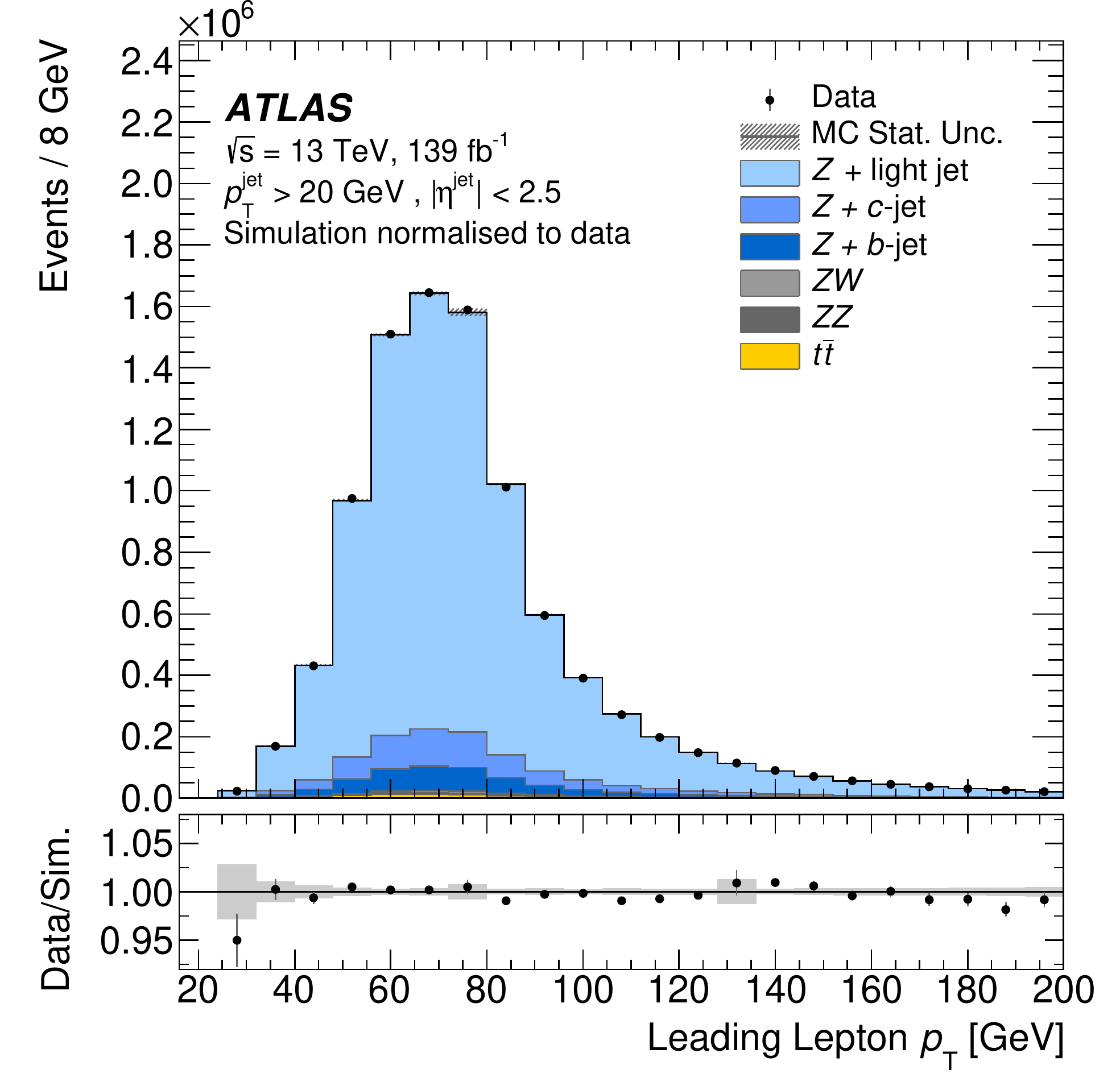}
\includegraphics[width=0.4\textwidth]{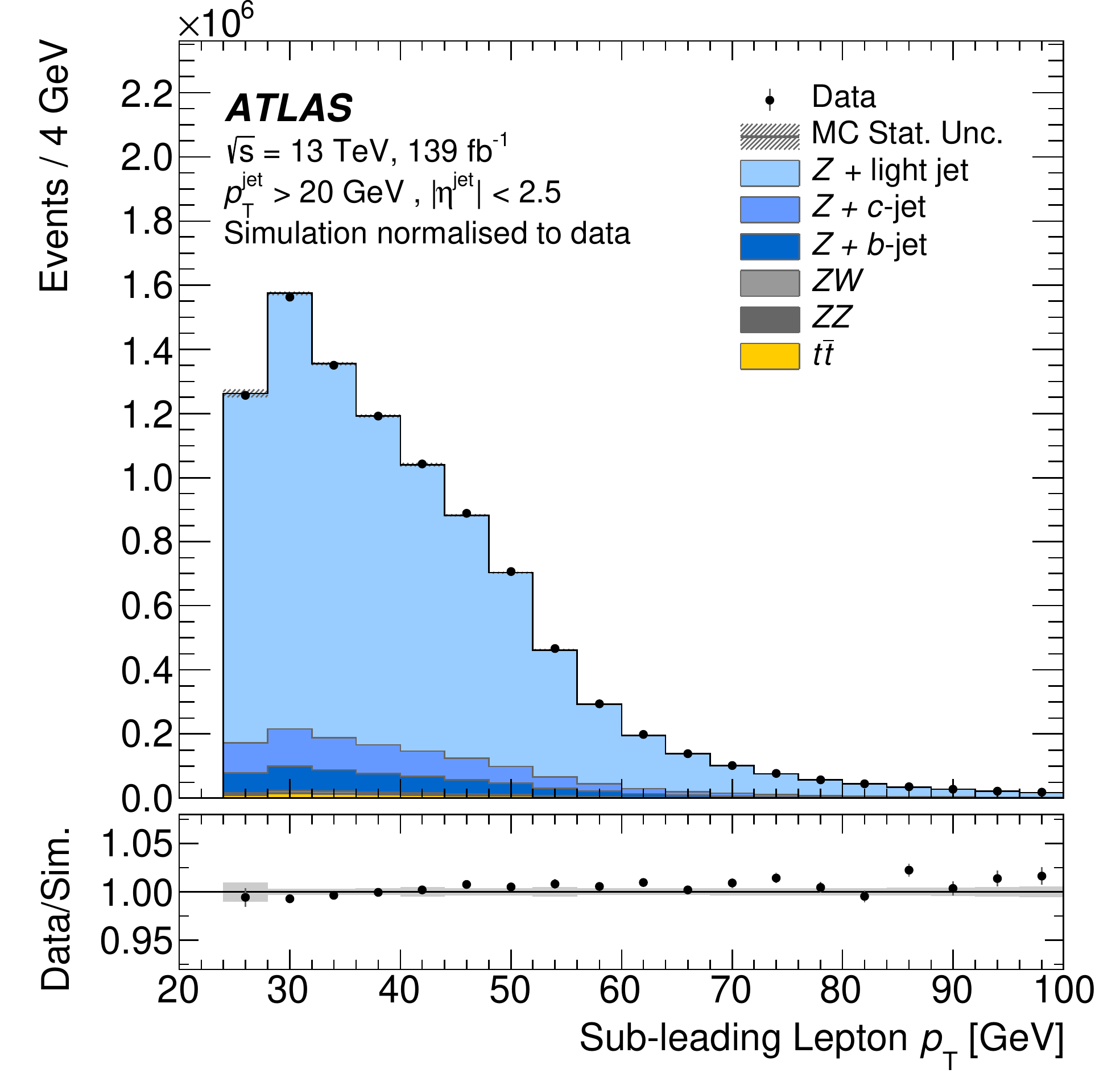}\\
\includegraphics[width=0.4\textwidth]{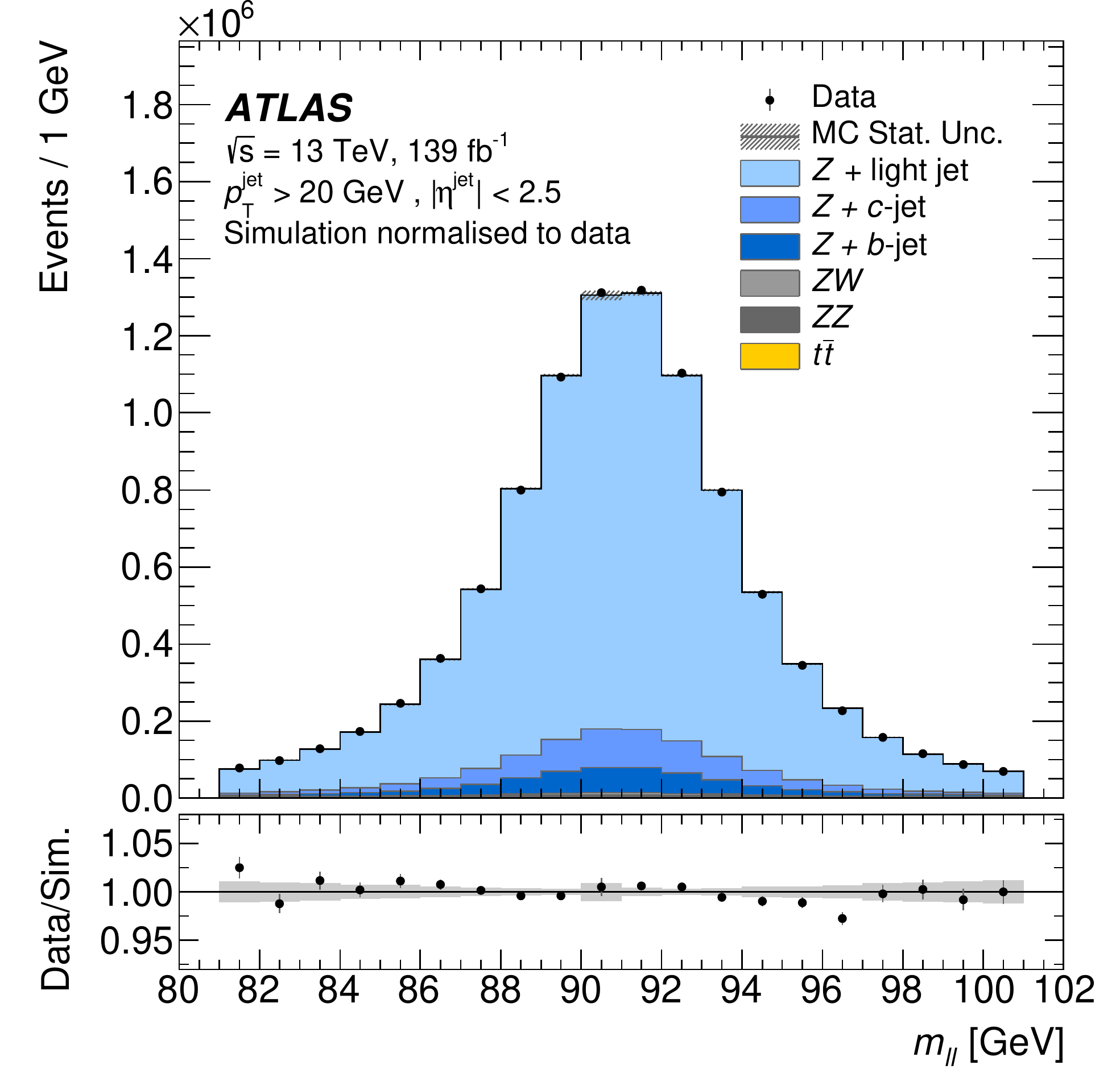}
\includegraphics[width=0.4\textwidth]{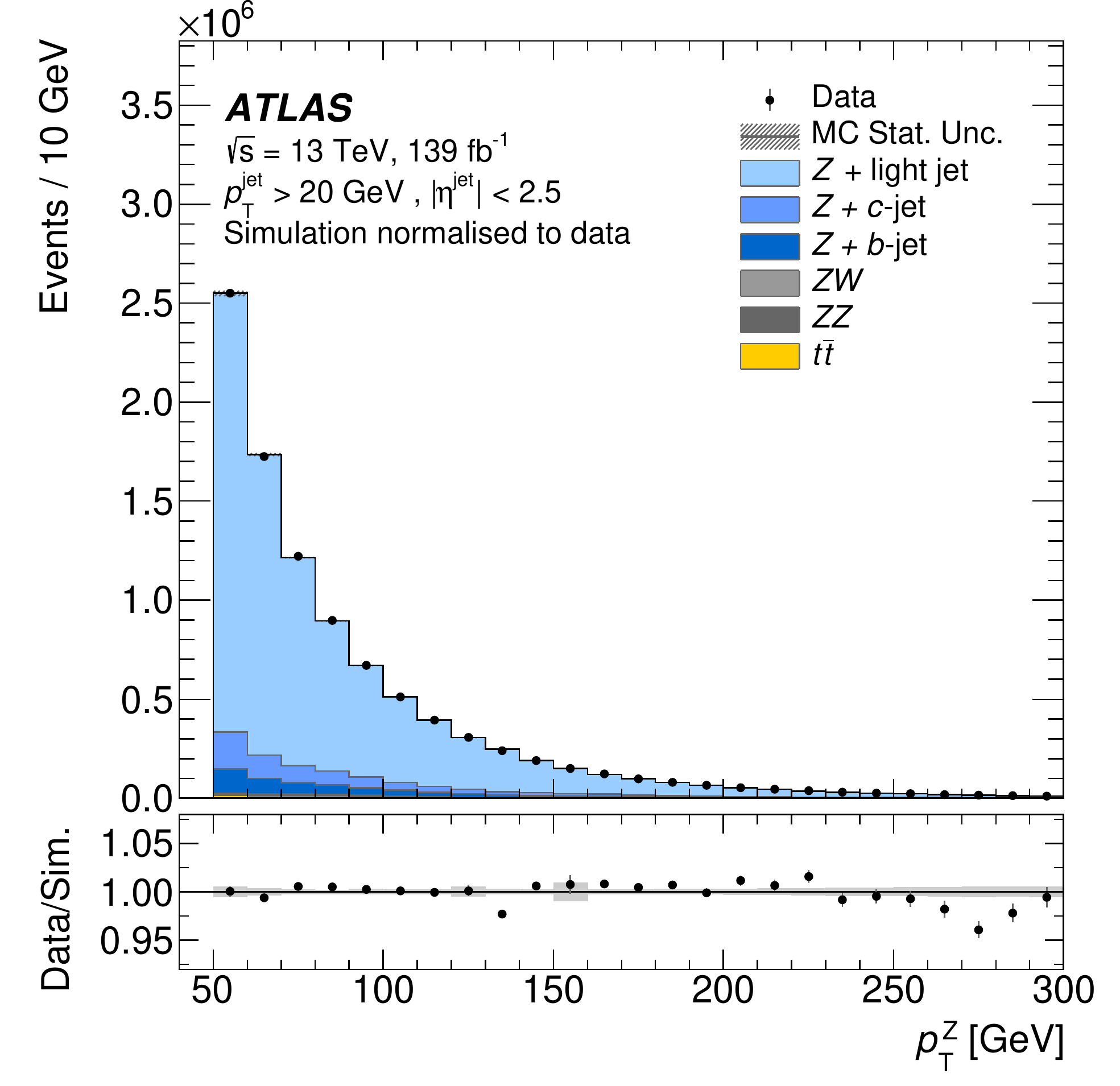}
\caption{Comparisons between data and simulation for the most relevant kinematic variables after event selection. The MC expectations are normalised to data. The shaded area corresponds to the statistical uncertainties of the MC simulations. The $Z$+jets events are simulated using \Sherpa[2.2.1]. Negative values of $m_{\mathrm{SV}}$ indicate that no SV was found in the jet.}
\label{fig:DataMC}
 
\end{figure}


\FloatBarrier
\section{The \text{DL1r} $b$-tagging algorithm}
\label{sec:btag}

Properties of $b$-hadrons, such as their relatively high mass, long lifetime and large multiplicity of reconstructed tracks from their decay products, are exploited to distinguish $b$-jets from $c$-jets and light-flavour jets.
Individual low-level taggers~\cite{ATL-PHYS-PUB-2016-012,ATL-PHYS-PUB-2017-003} are designed to target the specific properties of $b$-jets. The \texttt{IP2D}, \texttt{IP3D} and \texttt{RNNIP} algorithms exploit the individual properties of the tracks from charged particles matched to the jet, especially the track $d_{0}$ and $z_{0}$. \texttt{RNNIP} is based on a recurrent neural network and it explores the correlations between the IPs of different selected tracks in the jet.
The \texttt{SV1} algorithm exploits the output of the \texttt{SSVF} algorithm and \texttt{JetFitter}~\cite{ATL-PHYS-PUB-2018-025} reconstructs secondary and tertiary vertices, following the expected topology of $b$-hadron decays.
The \texttt{DL1r} algorithm is a deep neural network that takes the output of the low-level taggers and jet kinematics as input and builds a single discriminant~\cite{FTAG-2019-005,ATL-PHYS-PUB-2017-013}.

ATLAS analyses use selection requirements defining a lower bound on the \texttt{DL1r} discriminant to select $b$-jets with a certain efficiency.
Four of these so-called single-cut operating points (OPs) are defined, corresponding to $b$-jet selection efficiencies of 85\%, 77\%, 70\% and 60\%. The OPs are evaluated in a sample of $b$-jets from simulated $t\bar{t}$ events.
These selection requirements divide the \texttt{DL1r} score into five intervals to form the pseudo-continuous OPs, where the lower edge of the lowest interval corresponds to 100\% \texttt{DL1r} $b$-tagging efficiency, and the upper edge of the highest interval corresponds to 0\% efficiency.
 
The single-cut and pseudo-continuous OPs are calibrated in order to correct for efficiency differences between data and simulation.
The calibration SF, defined in Eq.~(\ref{eq:SF}), is measured relative to a reference efficiency $\varepsilon^{f}_{\mathrm{MC}}$.
The performance of the $b$-tagging algorithm in simulation is affected by the hadronisation and fragmentation model used in the parton shower simulation~\cite{ATLAS:2018sgt}.
To account for their differences, simulation-to-simulation SFs are applied to those simulated samples that have a fragmentation model different from the default.
The calculation of these simulation-to-simulation correction factors is described in Ref.~\cite{ATL-PHYS-PUB-2020-009}.
The top-quark pair production sample produced with \POWHEGBOX[v2]+~\Pythia[8.230] is used to define the reference MC efficiencies in the $b$- and $c$-jet calibrations.
As no equivalent $Z$+jets simulation produced with the \POWHEGBOX[v2] generator is available, the $Z$+jets \MADGRAPH+~\Pythia[8] simulation is used for the definition of the reference efficiency in the calculation of SF$^{\text{light}}$ for light-flavour jets.
Since the $Z$+jets \MADGRAPH+~\Pythia[8] simulation uses the same parton shower model and set of tuned parameters as the top-quark pair production simulation, the light-flavour jet results and the $b$- and $c$-jet calibrations are obtained relative to similar reference models.


\FloatBarrier

\section{The Negative Tag method}
\label{sec:method}

The \texttt{DL1r} algorithm rejects 97.52\% and 99.96\% of the light-flavour jets, respectively, at the 85\% and 60\% OPs, according to simulated $t\bar{t}$ events.
Therefore, the fraction of light-flavour jets passing the threshold is too low to estimate $\varepsilon^{\text{light}}$ in data.
The Negative Tag method~\cite{ATLAS-CONF-2014-046,D0:2006hcm} calibrates $\varepsilon^{\text{light}}_{\text{data}}$ using a modified version of the \texttt{DL1r} algorithm (called \texttt{DL1rFlip}) that achieves lower $\varepsilon^{b}$ to $\varepsilon^{\text{light}}$ ratios without changing $\varepsilon^{\text{light}}$ significantly.
 
Tracks matched to $b$-jets have relatively large and positively signed IPs due to the long lifetime of the $b$-hadrons and the presence of displaced decay vertices.
In contrast, tracks matched to light-flavour jets typically have IP values consistent with zero within the IP resolution such that a more symmetric\footnote{The tracks matched to light-flavour jets have a slight bias towards positive-sign values due to the presence of some long-lived particles (e.g.\ $K_\text{S}$ or $\Lambda$). The contribution from the mismodelling of long-lived particles is expected to be negligible relative to the mismodelling of the $d_0$ and $z_0$ resolutions~\cite{ATLAS-CONF-2018-006}.} IP distribution is expected.
The expected IP distributions of the tracks associated with $b$-jets, $c$-jets or light-flavour jets are shown in Figure~\ref{fig:signedIP}.
The Negative Tag method assumes that the probability for a light-flavour jet to be mistagged remains almost the same when inverting the IP signs of all tracks and displaced vertices.
This is based on the assumption that light-flavour jets are misidentified as $b$-jets mainly due to resolution effects in the track reconstruction which result in tracks and vertices from tracks with positive IPs inside the jet. Given the symmetric IP distributions, the fractions of tracks and vertices from tracks with positive IPs remain stable after inverting the IP signs of all tracks and vertices.
The presence of the positive tail in the IP distribution challenges this assumption and its impact is taken into account by a dedicated uncertainty, which is called the \texttt{DL1rFlip}-to-\texttt{DL1r} extrapolation uncertainty in the following.

\begin{figure}[!h]
\center
\includegraphics[width=0.48\textwidth ]{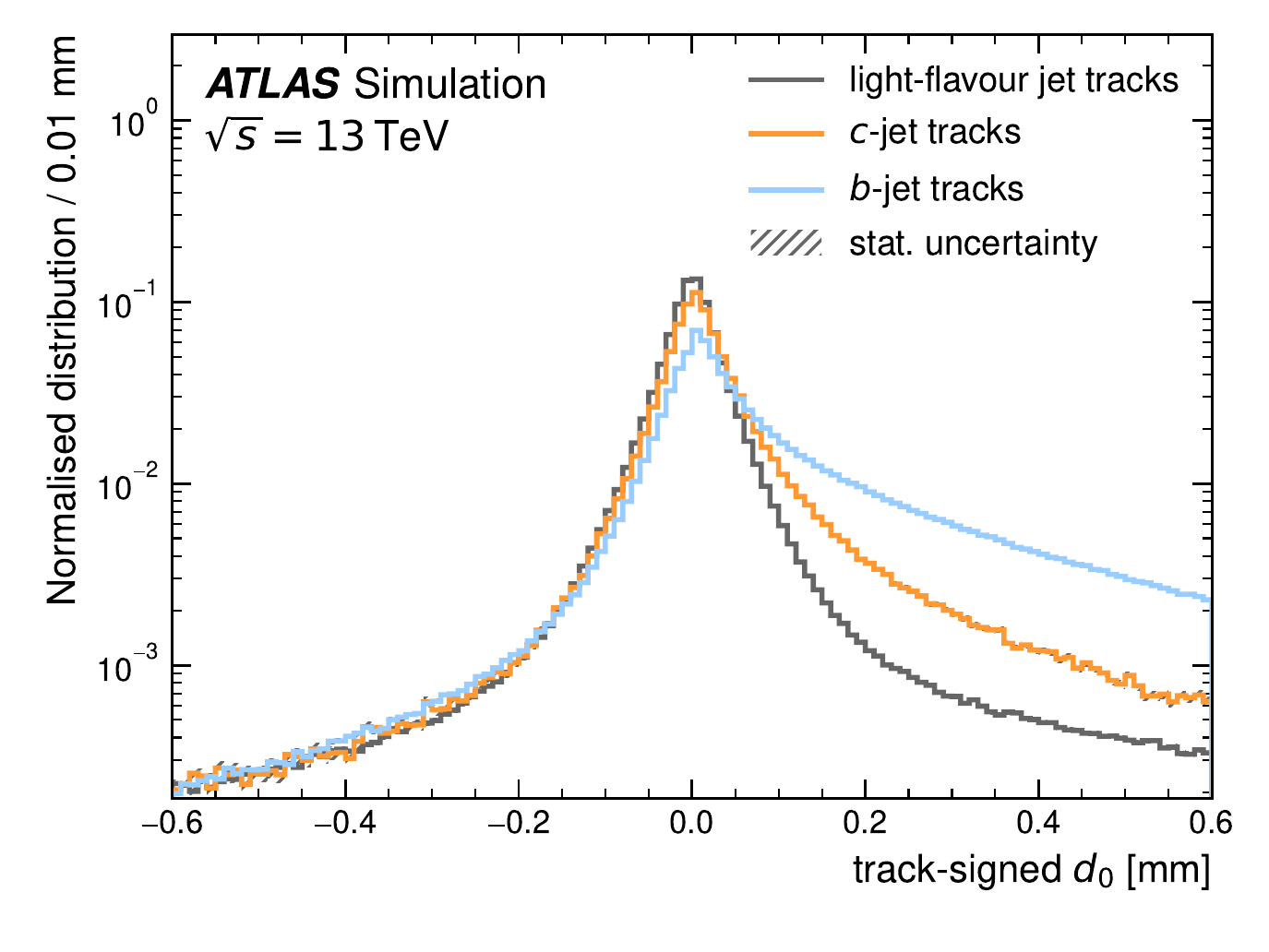}
\includegraphics[width=0.48\textwidth ]{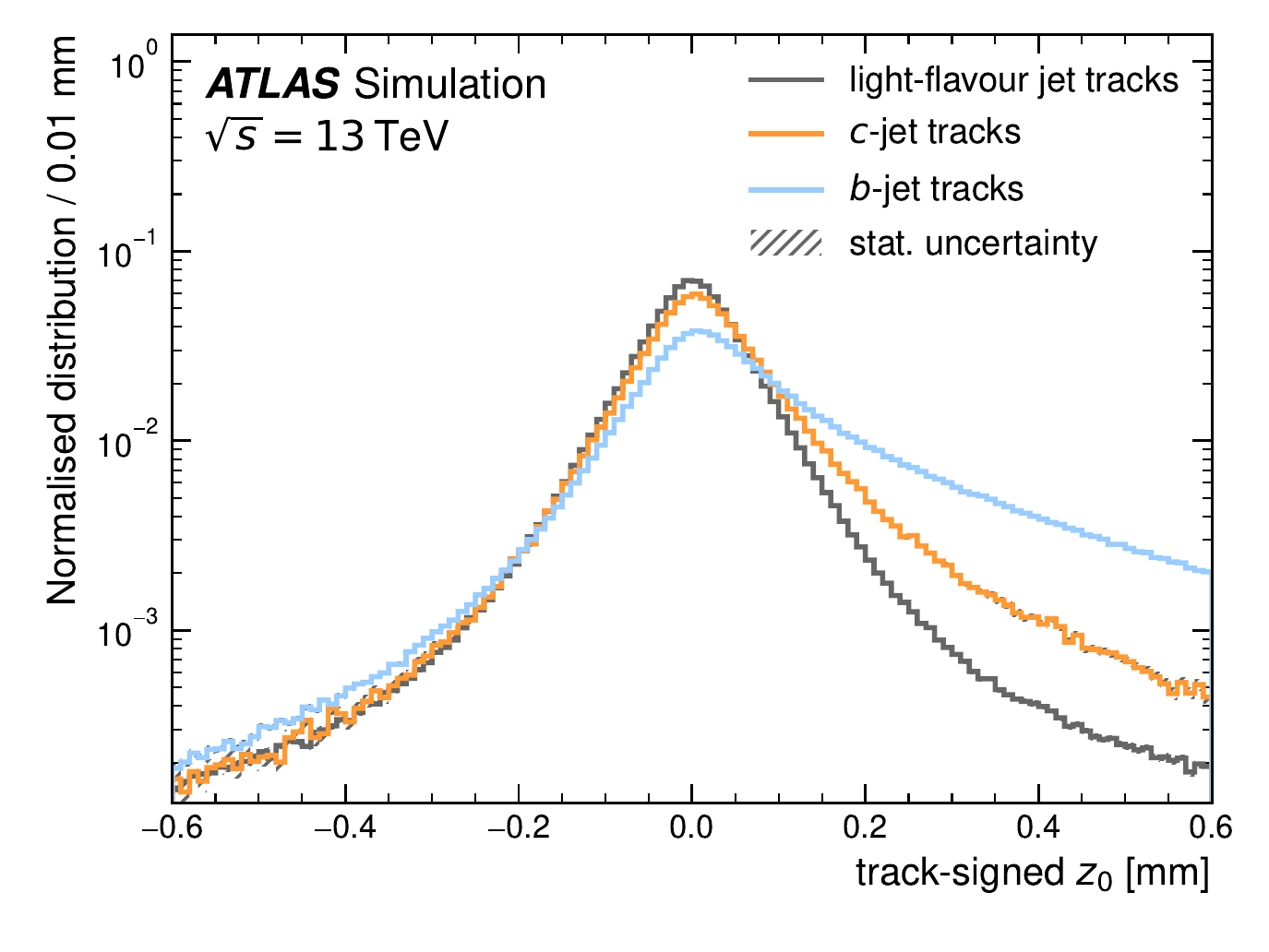}
\caption{
Signed-$d_{0}$ (left) and signed-$z_{0}$ (right) distributions for tracks matched to $b$-jets, $c$-jets and light-flavour jets in simulated $t\bar{t}$ events. The selected tracks are matched to particle-flow jets with $\pT> 20$~\GeV\ and $|\eta|< 2.5$, and pass the jet-vertex tagger selection. The distributions are normalised to unity. Statistical uncertainties are also shown.
}
\label{fig:signedIP}
\end{figure}
 
The \texttt{DL1rFlip} algorithm inverts the signs of the track IPs and the decay length\footnote{In this paper, vertices are assigned a negative decay length if they have an angle between the jet direction and the line joining the primary vertex to the secondary vertex of more than $\pi/2$ (see Section 14.2 of Reference~\cite{PERF-2012-04} for more details).},
while using the same algorithm training and OP-threshold definitions as the \texttt{DL1r} algorithm. The $\varepsilon^{\text{light}}$ values obtained by the two algorithms are approximately the same, while the \texttt{DL1rFlip} algorithm selects a smaller fraction of $b$-jets than the nominal version.
The discriminants of the \texttt{DL1r} and \texttt{DL1rFlip} algorithms are compared between data and simulation in Figure~\ref{fig:DL1rFlipVSNom}. These comparisons show that the heavy-flavour fractions are much lower for high \texttt{DL1rFlip} values than for high \texttt{DL1r} values. At the 85\% (60\%) single-cut OP, the \texttt{DL1r} algorithm selects a sample of jets in which 25\% (1\%) are light-flavour jets, while the \texttt{DL1rFlip} algorithm selects a sample in which a higher fraction, 38\% (6\%), are light-flavour jets at the same OP.
 
\begin{figure}[h!]
\centering
\includegraphics[width=0.49\textwidth]{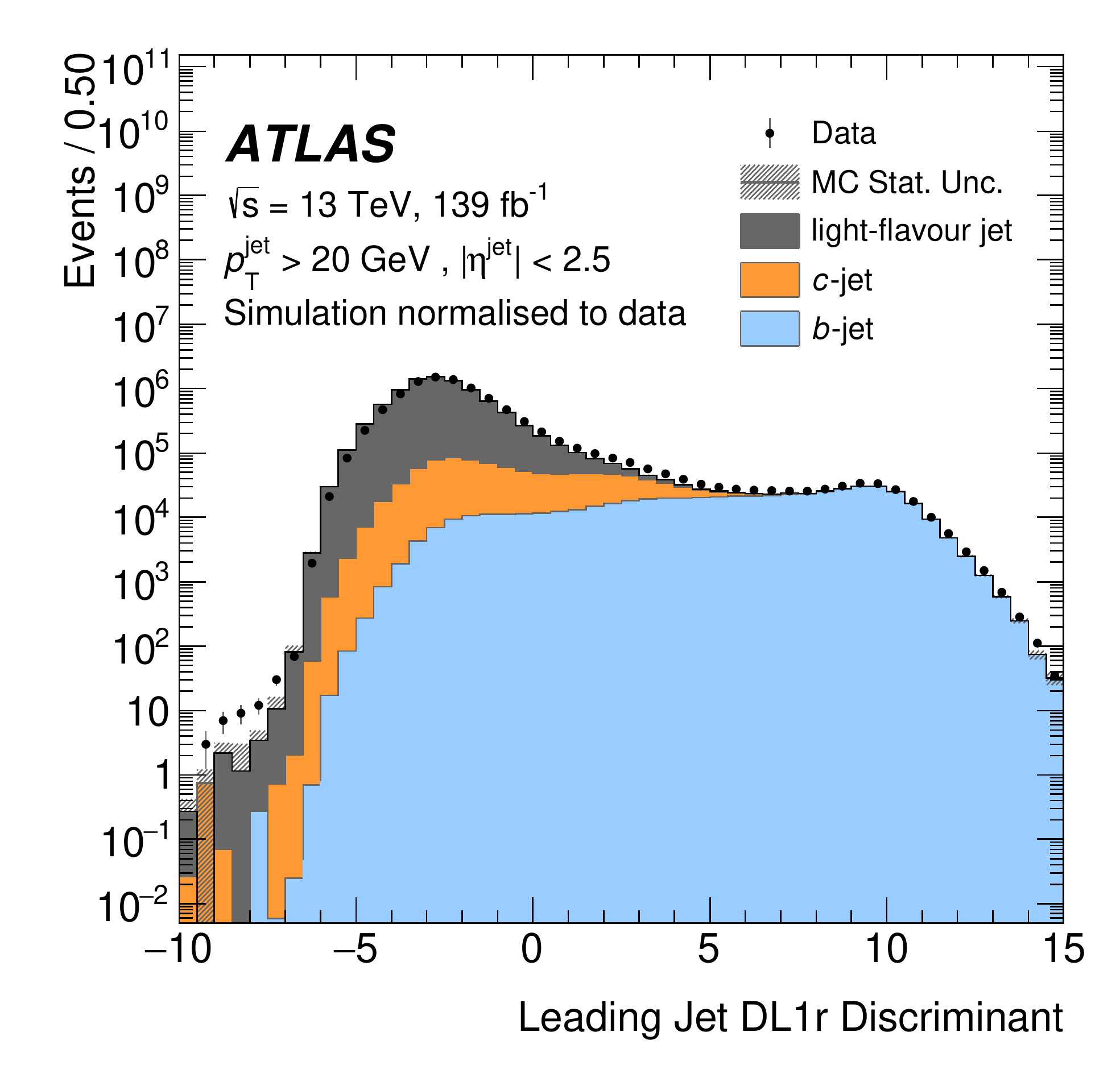}
\includegraphics[width=0.49\textwidth]{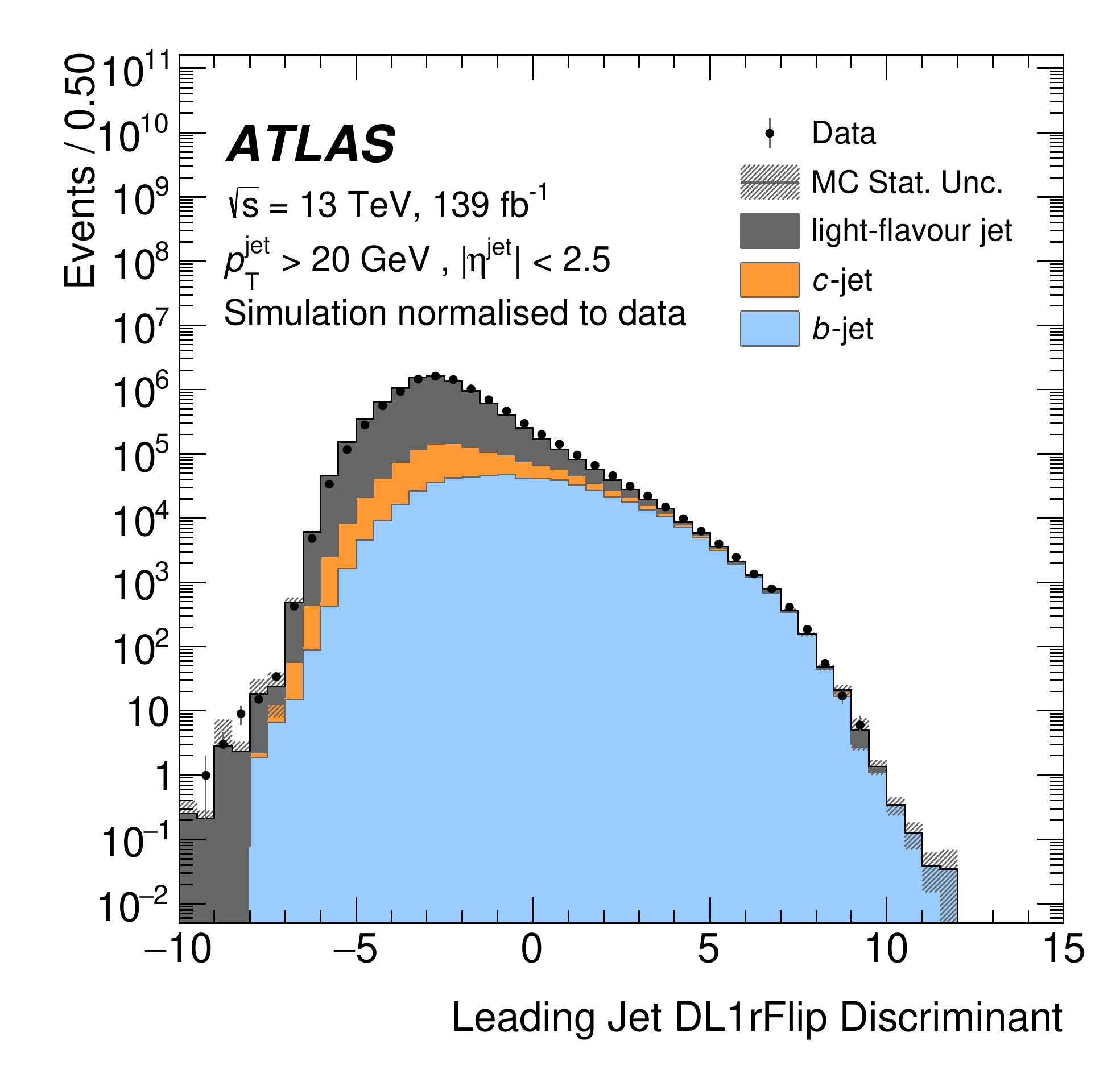}
\caption{Comparisons between data and simulation for the distributions of the \texttt{DL1r} discriminant (left) and \texttt{DL1rFlip} discriminant (right). Statistical uncertainties of the yields are shown.}
\label{fig:DL1rFlipVSNom}
\end{figure}
 
The calibration is performed independently in jet \pT intervals in order to account for the \pT dependence of $\varepsilon^{\text{light}}$.
A simultaneous binned fit to the $m_{\mathrm{SV}}$ distribution in each pseudo-continuous interval of the \texttt{DL1rFlip} discriminant is performed in order to simultaneously determine $\varepsilon^{b}_{\text{data}}$ and $\varepsilon^{\text{light}}_{\text{data}}$ in the \texttt{DL1rFlip} discriminant intervals.
The sensitivity of the fit does not allow the SFs of all three jet flavours to be derived simultaneously. Therefore, $\varepsilon^{c}_{\text{data}}$ is constrained to the MC predictions and SF$^{c}$ is fixed to unity within an uncertainty of 30\%, as suggested by studies of the $c$-jet mistagging efficiency calibration \cite{FTAG-2020-08}.

For a given interval of jet \pT, the expected number of jets for a defined discriminant interval $i$ is given by
 
\begin{equation*}
N_{i}(m_{\text{SV}}) = N_{i,\text{MC}} \times C \times \left[ \sum_{f=\text{light},c,b} F^{f} \times \text{SF}^{f}_{i} \times \varepsilon^{f}_{i, \text{MC}} \times P^{f}_{i}\!(m_{\text{SV}}) \right ],
\end{equation*}
 
where $N_{i,\text{MC}}$ is the predicted flavour-inclusive event yield for each discriminant interval; $C$ is a global normalisation factor and $F^{f}$ are the jet-flavour fractions; $P_{i}^{f}\!(m_{\text{SV}})$ is the probability density function of $m_{\text{SV}}$ for jet flavour $f$ in the $i$-th \texttt{DL1rFlip} discriminant interval, taken from simulation. The $P_{i}^{f}\!(m_{\text{SV}})$ is defined in such a way to integrate an additional bin ($m_{\text{SV}}<0$~\GeV) representing the number of events where no secondary vertex is found. The $m_{\text{SV}}$ has been obtained with tracks with nominal sign as input to the \texttt{SSVF} algorithm.
 
The $C$, $F^{f}$ and $\text{SF}^{b,\text{light}}_{i}$ parameters are allowed to float in the fit, while $N_{i,\text{MC}}$ and $\varepsilon^{f}_{i,\text{MC}}$ are fixed to the predictions from simulated events and $\text{SF}^{c}_{i}$ is set to $1.0\pm 0.3$. The constraints $\sum_{i=1}^{5} \varepsilon_{i, \text{MC}}^{f} \times \text{SF}_{i}^{f} = 1$, where the contributions to the first bin (corresponding to the 100\%$-$85\% OP interval) are allowed to vary such that unitarity is preserved, and $F^{\text{light}} = 1 - F^{b} - F^{c}$ are applied, reducing the number of free parameters to 11. Figure \ref{fig:2Dfit} shows the post-fit $m_{\mathrm{SV}}$ and \texttt{DL1rFlip} discriminant distributions for the 50--100~\GeV\ \pT interval. The parameters of interest, i.e.\ the $\text{SF}^{\text{light}}_{i}$, are obtained from the \texttt{DL1rFlip} algorithm. In order to apply this calibration to the \texttt{DL1r} algorithm, an extrapolation uncertainty is added to account for the differences between these two tagging algorithms. Details of this \texttt{DL1rFlip}-to-\texttt{DL1r} extrapolation uncertainty are presented in Section~\ref{sec:unc}.
 
\begin{figure}[tbh!]
\centering
\begin{subfigure}[t]{0.48\textwidth}
\includegraphics[width=\textwidth]{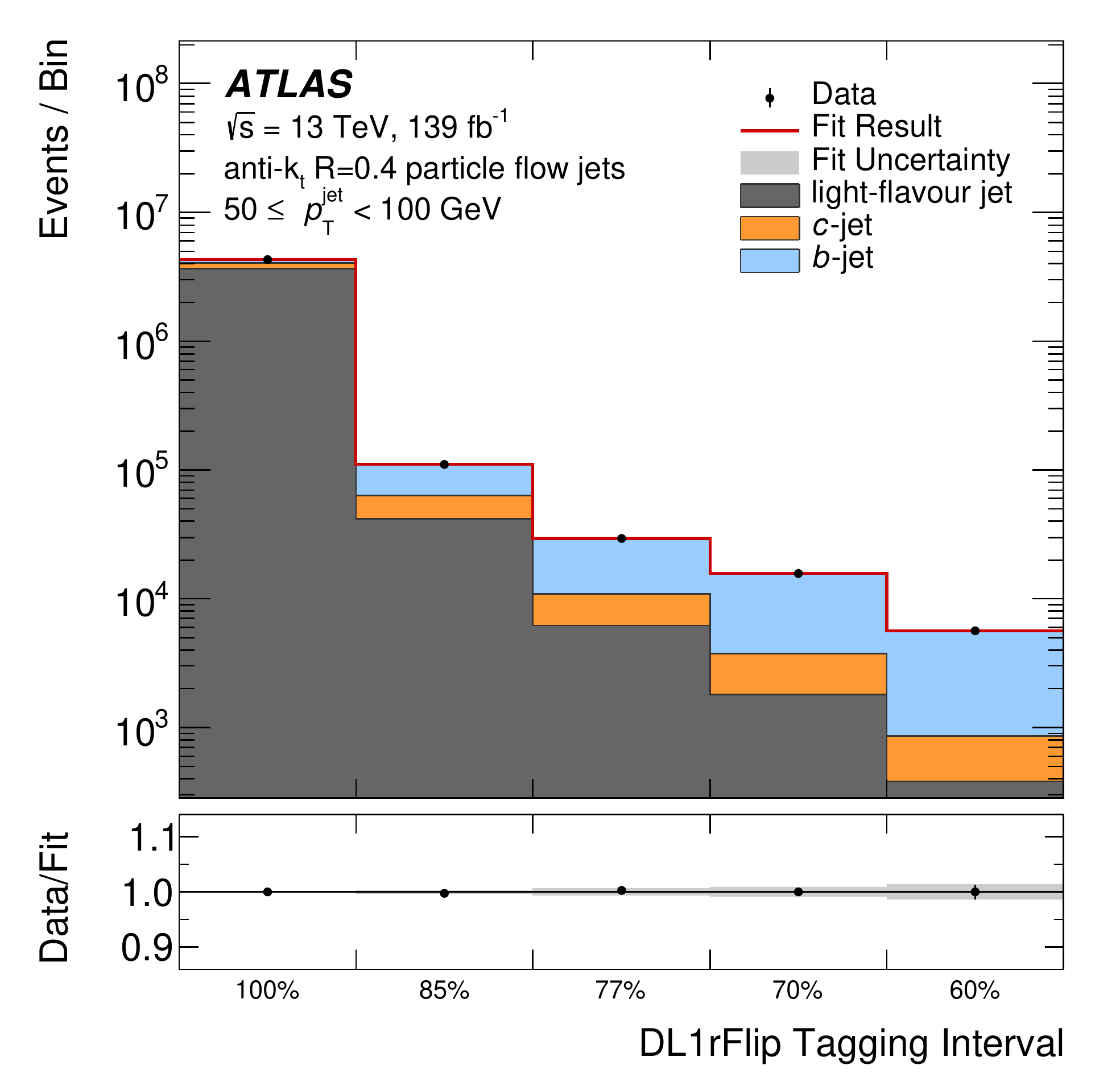}
\end{subfigure}
\begin{subfigure}[t]{0.48\textwidth}
\includegraphics[width=\textwidth]{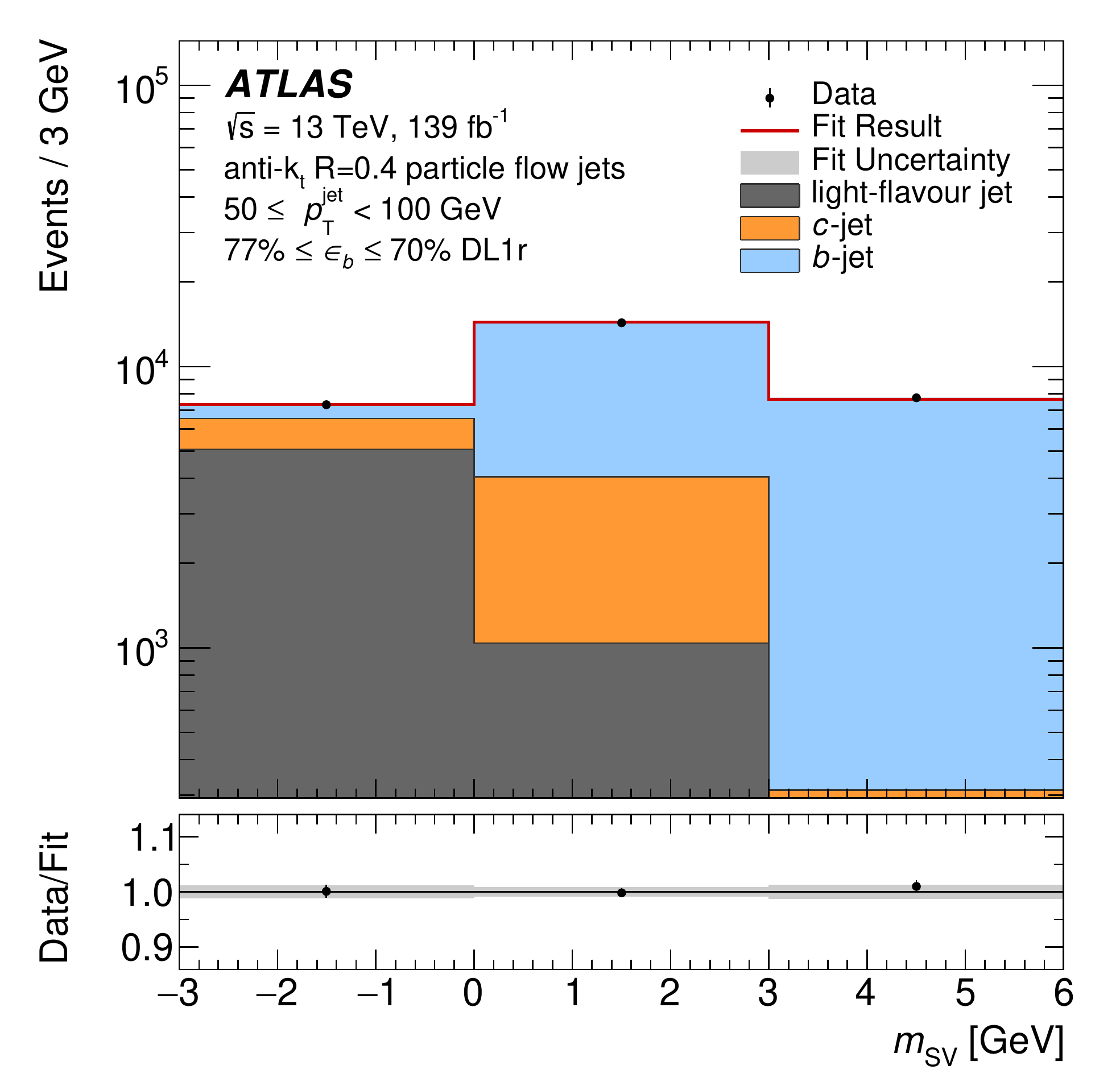}
\end{subfigure}
\caption{Post-fit distribution of the \texttt{DL1rFlip} discriminant (left) and the $m_{\mathrm{SV}}$ distribution in the 77\%--70\% OP interval (right). Both distributions are shown for the 50--100~\GeV\ \pT interval. The shaded area corresponds to the total statistical uncertainty. The binning in $m_{\mathrm{SV}}$ allows good separation between the jet flavours and also uses the minimum number of bins necessary to constrain the free parameters of the fit. Choosing the coarsest binning possible allows the statistical uncertainties to be minimised. }
\label{fig:2Dfit}
\end{figure}


\FloatBarrier
 
\section{Systematic uncertainties}
\label{sec:unc}

The measurement of SF$^{\text{light}}$ is affected by four types of uncertainties, including those due to experimental effects, the modelling of the $Z$+jets and background processes, and the limited number of events in data and simulation.
For each source of uncertainty, one parameter of the fit model is varied at a time, and the effect of this variation on SF$^{\text{light}}$ is evaluated. This approach is chosen to prevent the fit from using the data to constrain the impact of individual systematic uncertainties.
The probability density functions in $m_{\mathrm{SV}}$ are derived from the \Sherpa~2.2.1 $Z$+jets simulation sample. The probability density function for each flavour component is separately normalised to unity, so only the shape of the $m_{\mathrm{SV}}$ distributions is estimated using the \Sherpa~2.2.1 $Z$+jets simulation. The simulated efficiencies from the \MADGRAPH $Z$+jets simulation listed in Table~\ref{tab:mc} are used in the likelihood definition to predict the expected number of events for each flavour category. Given that the flavour fractions and most SFs are extracted from the data, the systematic uncertainty calculation considers mainly effects on the template shapes.
An additional uncertainty accounts for differences between SF$^{\text{light}}$ values from the \texttt{DL1rFlip} and \texttt{DL1r} algorithms to ensure the applicability of this calibration to the \texttt{DL1r} algorithm.
 
Uncertainties in the modelling of the detector response have a negligible impact on the calibration results.
The impact of the jet energy scale uncertainties was estimated by globally shifting the jet energy scale (JES) by 5\%, in accord with the most conservative estimate of the JES uncertainty in \pT derived in Ref.~\cite{JETM-2018-05}. This conservative assumption about the magnitude of the jet energy scale uncertainties leads to an estimated impact of less than 1\% on the calibration results, owing to the low \pT dependence of the scale factors.
The impact of $d_{0}$ and $z_{0}$ IP resolution modelling uncertainties on the $m_{\mathrm{SV}}$ templates was estimated by applying data-driven corrections to the simulated $d_{0}$ and $z_{0}$ in a sample of simulated top-quark pairs with a method similar to the one in Ref.~\cite{ATLAS-CONF-2018-006}. The impact of these corrections is transfered to the fit templates and is symmetrised around the central value.
This approach was chosen in order to mitigate the effect of statistical fluctuations in the available $Z$+jets samples.
Other experimental uncertainties were found to be negligible because they are not correlated with the discriminant used in the fit. 
 
Simulations of $Z$+jets via \Sherpa[2.2.1] are used as the nominal model to derive the fit templates for $b$-jets, $c$-jets and light-flavour jets in the $m_{\mathrm{SV}}$ distribution.
In order to assess the impact of the resulting choice of parton shower model, shower matching scheme and order of the perturbative QCD calculations on the template shapes, SFs$^{\text{light}}$ are derived using an alternative model, provided by the $Z$+jets \MADGRAPH+~\Pythia[8] simulation.
The difference between the obtained SFs$^{\text{light}}$ is taken as the estimate of the MC modelling uncertainty.
The effect of QCD scale uncertainties is estimated by independently doubling or halving the renormalisation and factorisation scales. The impact of choosing a particular PDF set is estimated by propagating its uncertainties.
However, the impacts of the QCD scale and PDF set uncertainties are found to be covered by the statistical uncertainties of the MC simulation and these uncertainties are therefore not included.

The effect of the 30\% uncertainty in SF$^{c}$, which was fixed to a value of 1.0 in the fit, is estimated by repeating the fit with SF$^{c}$ set to 1.3 and 0.7 instead.
The mean impact of the two variations on SF$^{\text{light}}$ is taken as the impact of the charm calibration uncertainty on the light-flavour jet calibration.

The $\varepsilon^{\text{light}}$ value is assumed to be the same for \texttt{DL1rFlip} and \texttt{DL1r} to first order, as described in Section~\ref{sec:method}.
However, mismodelling of the IP resolution, the fake-track rate or the parton shower can have different effects on the tagging performance and SF$^{\text{light}}$ values of the \texttt{DL1r} and \texttt{DL1rFlip} algorithms.
A \texttt{DL1rFlip}-to-\texttt{DL1r} extrapolation uncertainty is added to account for residual differences in SF$^{\text{light}}$ between the two algorithms.
SF$^{\text{light}}$ cannot be measured in data for the \texttt{DL1r} algorithm. Therefore, a method similar to the one in Ref.~\cite{ATLAS-CONF-2018-006} is used to derive SF$^{\text{light}}$ for \texttt{DL1rFlip} and \texttt{DL1r} from MC simulation in order to compare the two calibrations.
The \texttt{DL1rFlip}-to-\texttt{DL1r} extrapolation uncertainty consists of two components. One component estimates the impact of the modelling of tracking variables on the \texttt{DL1rFlip} and \texttt{DL1r} tagging performance, evaluated in a bottom-up approach correcting the simulation, and the second component estimates the impact of the shower and hadronisation model on the tagging performance difference between \texttt{DL1rFlip} and \texttt{DL1r}.
The impact of the modelling of tracking variables is evaluated by applying data-driven corrections to underlying tracking variables affecting the $b$-tagging performance in a sample of simulated \ttbar events generated with \POWHEGBOX[v2] +~\Pythia[8.230].
The impact on $\varepsilon^{\text{light}}$ of correcting these observables is evaluated with respect to the uncorrected simulation.
Only corrections which have been shown to have a significant impact are considered~\cite{ATLAS-CONF-2018-006}. These include the IP resolution and fake-track rate modelling~\cite{ATL-PHYS-PUB-2015-051}.
The total impact on the mistagging efficiency is obtained by multiplying the effects of all corrections.
The total impact on $\varepsilon^{\text{light}}_{\text{MC}}$ is compared between \texttt{DL1r} and \texttt{DL1rFlip} and the relative difference is assigned as the first component of the \texttt{DL1rFlip}-to-\texttt{DL1r} extrapolation uncertainty.
A comparison of these simulation-based calibration factors is shown in Figure~\ref{fig:AdjustedMCcalib}.
\begin{figure}[h!]
\centering
\includegraphics[width=0.49\textwidth]{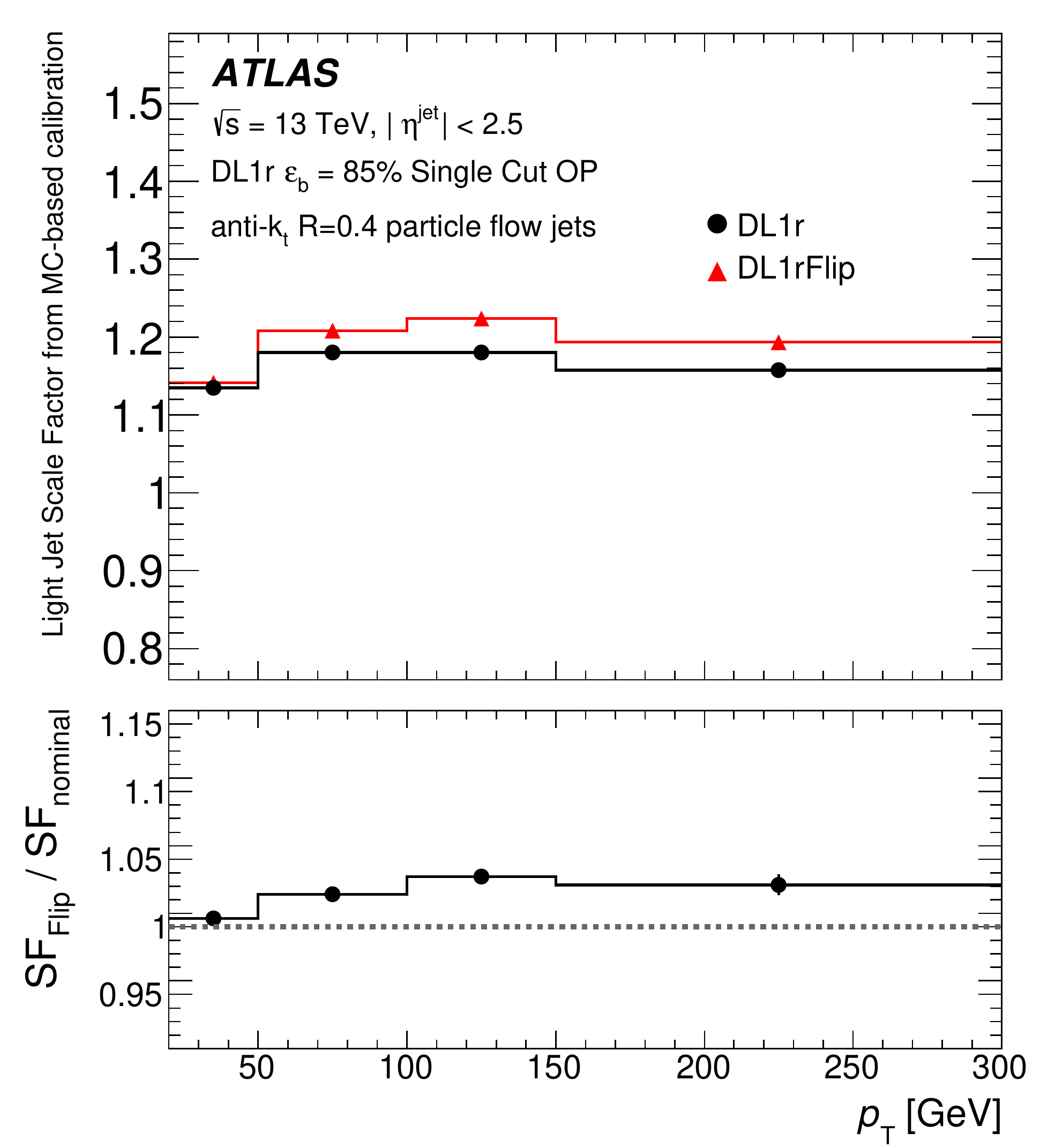}
\includegraphics[width=0.49\textwidth]{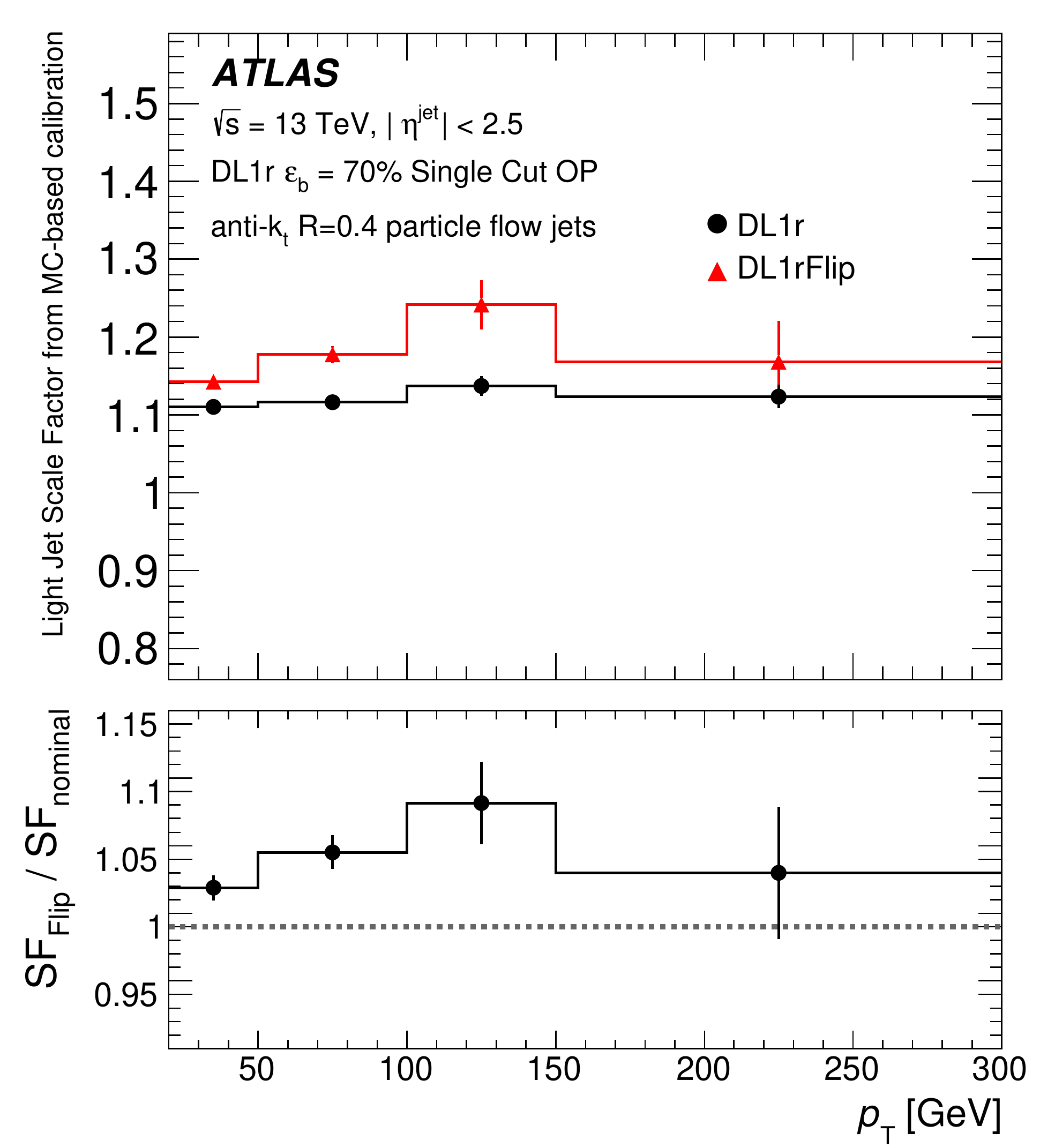}
\caption{Comparison of the light-flavour jet efficiency scale factor results for the \texttt{DL1r} and \texttt{DL1rFlip} tagger obtained with the MC-based calibration on a \POWHEGBOX[v2] +~\Pythia[8.230] \ttbar MC simulation, considering the 85\% and 70\% operating points, as a function of the jet \pT. The error bars correspond to MC statistical uncertainties. The lower panel shows the ratio of \texttt{DL1rFlip} to \texttt{DL1r} tagger MC-based calibration efficiency scale factors.
The envelope of this ratio across the considered jet \pT interval is used as an estimate for the impact of the tracking variable mismodelling on the difference between the \texttt{DL1rFlip} and \texttt{DL1r} calibrations, which is a part of the \texttt{DL1rFlip}-to-\texttt{DL1r} extrapolation uncertainty, estimating the impact of applying the calibrations derived for the \texttt{DL1rFlip} algorithm to the \texttt{DL1r} algorithm.
}
\label{fig:AdjustedMCcalib}
\end{figure}
The impact of the parton shower modelling on the difference between SF$^{\text{light}}$ in \texttt{DL1r} and \texttt{DL1rFlip} is estimated by multiplying the SFs$^{\text{light}}$ for \texttt{DL1r} (\texttt{DL1rFlip}), derived using the corrected \POWHEGBOX[v2] +~\Pythia[8.230] simulation, by the simulation-to-simulation SFs defined as the ratio of the \texttt{DL1r} (\texttt{DL1rFlip}) $\varepsilon^{\text{light}}$ for the \Sherpa[2.2.1] and the \MADGRAPH+~\Pythia[8] simulations. The simulation-to-simulation SFs have been derived using the $Z$+jets samples listed in Table~\ref{tab:mc}.
This gives an estimate of the \texttt{DL1rFlip}-to-\texttt{DL1r} extrapolation uncertainty for an alternative shower model. The difference between this estimate and the estimate obtained using the corrected \POWHEGBOX[v2] +~\Pythia[8.230] simulation is used as the component assessing the impact of the shower and hadronisation model on the tagging performance difference between \texttt{DL1rFlip} and \texttt{DL1r}.
The two components are added in quadrature to obtain the \texttt{DL1rFlip}-to-\texttt{DL1r} extrapolation uncertainty.
The envelope over \pT of the \texttt{DL1rFlip}-to-\texttt{DL1r} extrapolation uncertainty is used for the final uncertainty estimate.
The shower modelling makes the largest contribution to the \texttt{DL1rFlip}-to-\texttt{DL1r} extrapolation uncertainty and ranges up to 10\%. The largest uncertainty contribution in correcting the MC simulation is obtained from the track IP resolution modelling corrections, ranging up to 10\% as well.
The total \texttt{DL1rFlip}-to-\texttt{DL1r} extrapolation uncertainty is 10\%--12\% depending on the $b$-tagging OP and is the dominant systematic uncertainty overall.
 
Statistical uncertainties of the data and the $m_{\mathrm{SV}}$ templates from simulated MC events are taken into account.
The statistical uncertainties of the MC-based fit templates are implemented following the light-weight Beeston--Barlow method~\cite{Cranmer:1456844}.
As their effect is non-negligible, cross-check studies were performed using a `toy' approach. The fit template bin entries were fluctuated randomly around the nominal value according to a Gaussian probability distribution with a standard deviation equal to the MC statistical uncertainty in the template bins. SF$^{\text{light}}$ was extracted for each of these new fit templates and the standard deviation of the resulting SF$^{\text{light}}$ distribution was taken as the statistical uncertainty. The statistical uncertainty estimates via the Beeston--Barlow model and the toy approach were found to be consistent.


\FloatBarrier

\section{Results}
\label{sec:result}

The calibration SFs$^{\text{light}}$ of the \texttt{DL1r} algorithm are presented. The 85\%, 77\% and 70\% OPs were successfully calibrated in data by using the Negative Tag method. However, it is not feasible to calibrate the 60\% OP in data because of insufficient statistics and the relatively large contamination by heavy-flavour jets.
 
Due to these limitations and since the measured SFs for the closest 77\%--70\% and 70\%--60\% pseudo-continuous OPs do not show any significant deviation from unity compared to their respective measurement uncertainties, the assumption is made that the same must hold for the 60\% OP. The SF for this OP is thus set to 1. Its dominant uncertainty, the \texttt{DL1rFlip}-to-\texttt{DL1r} extrapolation uncertainty, is derived specifically for the 60\% OP, and found to be slightly larger than for the 70\% OP. The other (sub-leading) uncertainties are assumed to be identical to the 70\% OP, since their evaluation suffers from large uncertainties. It should also be noted that extracting very precise SFs$^\text{light}$ for the 60\%--0\% pseudo-continuous OP interval is not critical to most physics analyses as only around 0.5~per~mille of all light-flavour jets are $b$-tagged at this OP~\cite{FTAG-2019-005}.

The results for the mistagging efficiency calibration of light-flavour jets from the \texttt{DL1r} tagger, using particle-flow jets in Run 2 data, are shown in Figure \ref{fig:DL1rPFlowRun2Results_PC} for the pseudo-continuous OPs of the \texttt{DL1r} algorithm. The calibration of the \texttt{DL1r} single-cut OPs, which is derived from the calibration results of the pseudo-continuous OPs, is shown in Figure \ref{fig:DL1rPFlowRun2Results}. Furthermore, Figure~\ref{fig:DL1rPFlowRun2Uncertainties} shows a breakdown of the uncertainties in the SFs$^{\text{light}}$ for each jet \pT interval and single-cut OP.
 
The measured SFs$^{\text{light}}$ are consistent with unity within uncertainties, except for the 85\%--77\% pseudo-continuous OP and the 85\% single-cut OP, where the data prefer SFs$^{\text{light}}$ which differ from one by about one standard deviation.
Therefore, the results indicate the MC simulation predicts $\varepsilon^{\text{light}}$ to be similar to that in data.
 
Mistags of light-flavour jets are mostly caused by the presence of fake tracks and the limited impact parameter resolution~\cite{ATLAS-CONF-2018-006}. The fake-track rate is underestimated by the MC simulation and the impact parameter resolution is overestimated~\cite{ATL-PHYS-PUB-2015-051,ATLAS-CONF-2018-006}. An increase in the fake-track rate and a decrease in impact parameter resolution both give rise to larger mistagging efficiencies~\cite{ATLAS:2009zsq,ATLAS-CONF-2018-006}. Observed SFs slightly larger than unity are therefore expected.

Tables~\ref{tab:CaloDL1r_85}--\ref{tab:CaloDL1r_60} feature a detailed breakdown of the uncertainties in the results.
Overall, the largest systematic uncertainty is the \texttt{DL1rFlip}-to-\texttt{DL1r} extrapolation uncertainty, with values of 10\%--12\%, depending on the $b$-tagging OP.
The impact of the charm calibration uncertainty amounts to 5\%--10\% depending on the jet \pT and the $b$-tagging OP, and is therefore one of the larger uncertainties.
The MC modelling uncertainty is a few percent in most cases, but reaches its maximum of 11\% for the 70\% single-cut OP in the $20 \leq \text{jet}~\pT \leq 50$~\GeV\ interval.
 
The effect of the $d_{0}$ and $z_{0}$ impact parameter resolution modelling uncertainty is generally of the order of a few percent and therefore mostly subdominant. However, the effect can range up to 14\% for low jet \pT and tight OPs.
The data statistical uncertainties give subdominant contributions except for high jet \pT and tight OPs, where contributions range from 0.5\% to 9.5\%.
The MC statistical uncertainty is one of the dominant contributions, ranging from 0.4\% to 13\%, depending on the OP and the jet \pT.
The total uncertainty in SF$^{\text{light}}$ is between 11\% and 23\%.
In general, higher precision is obtained for the looser OPs and the size of the dominant uncertainties does not depend significantly on the jet \pT.
 
\begin{figure}[h!]
\centering
\includegraphics[width=0.49\textwidth]{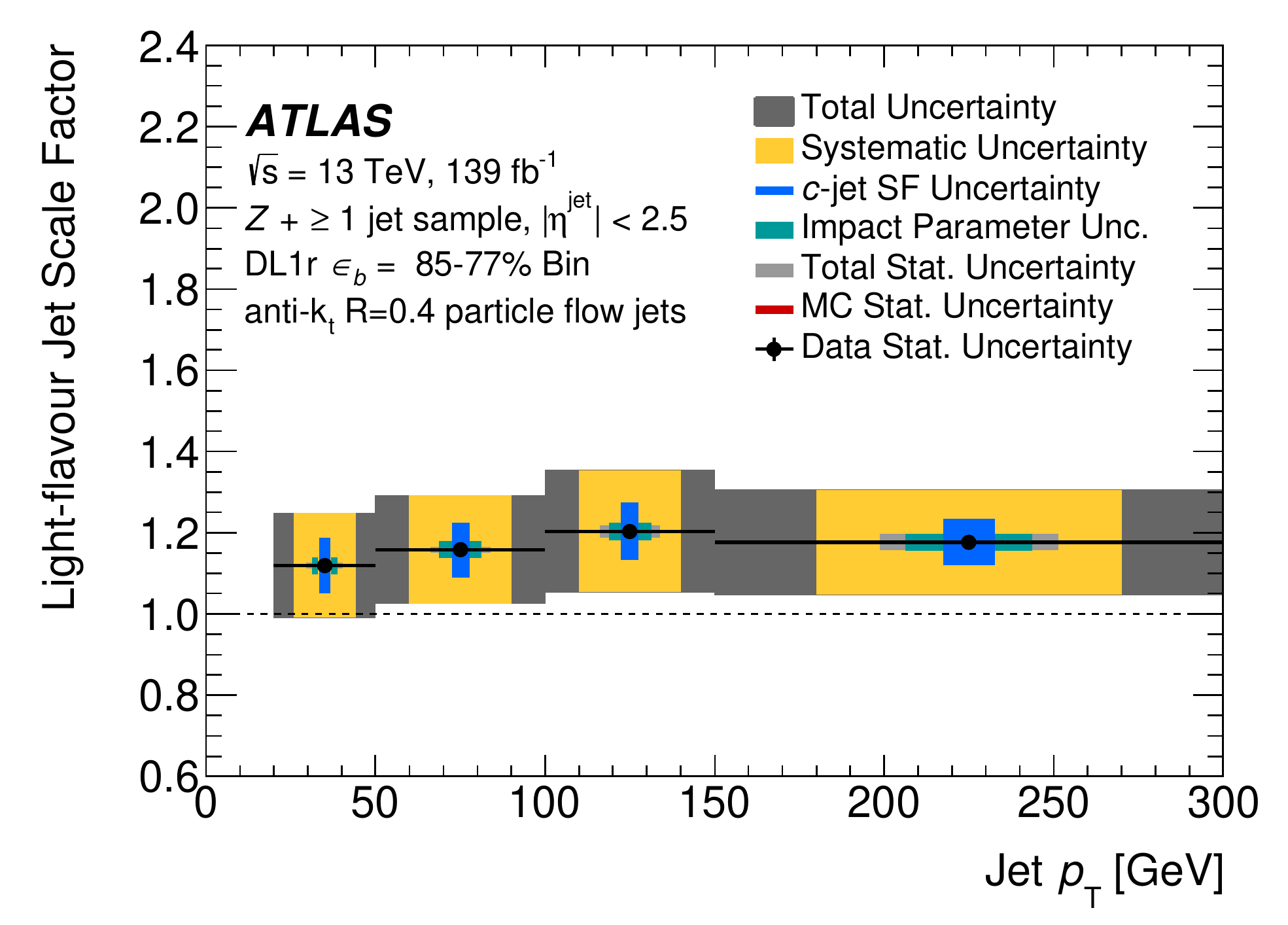}
\includegraphics[width=0.49\textwidth]{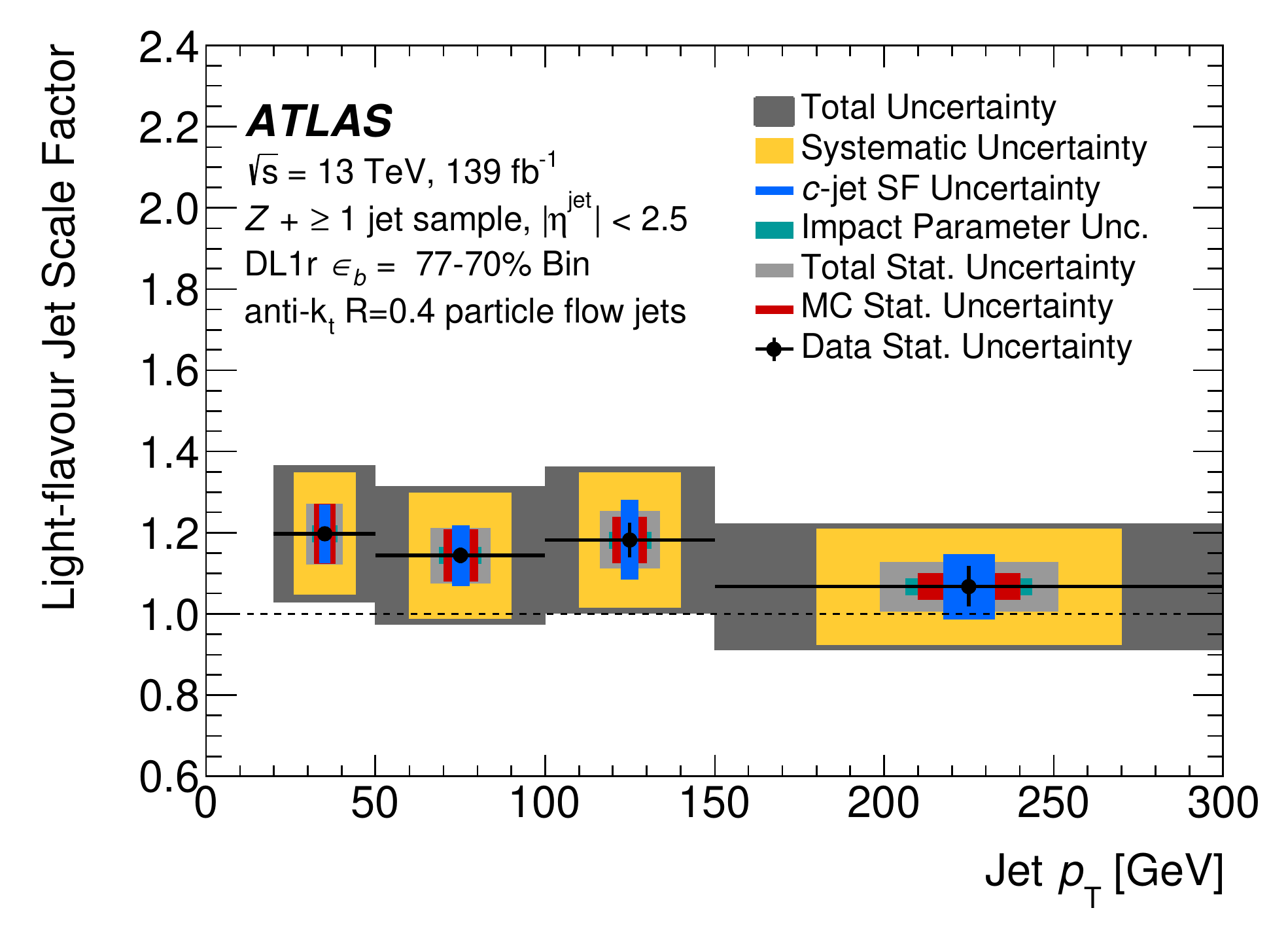} \\
\includegraphics[width=0.49\textwidth]{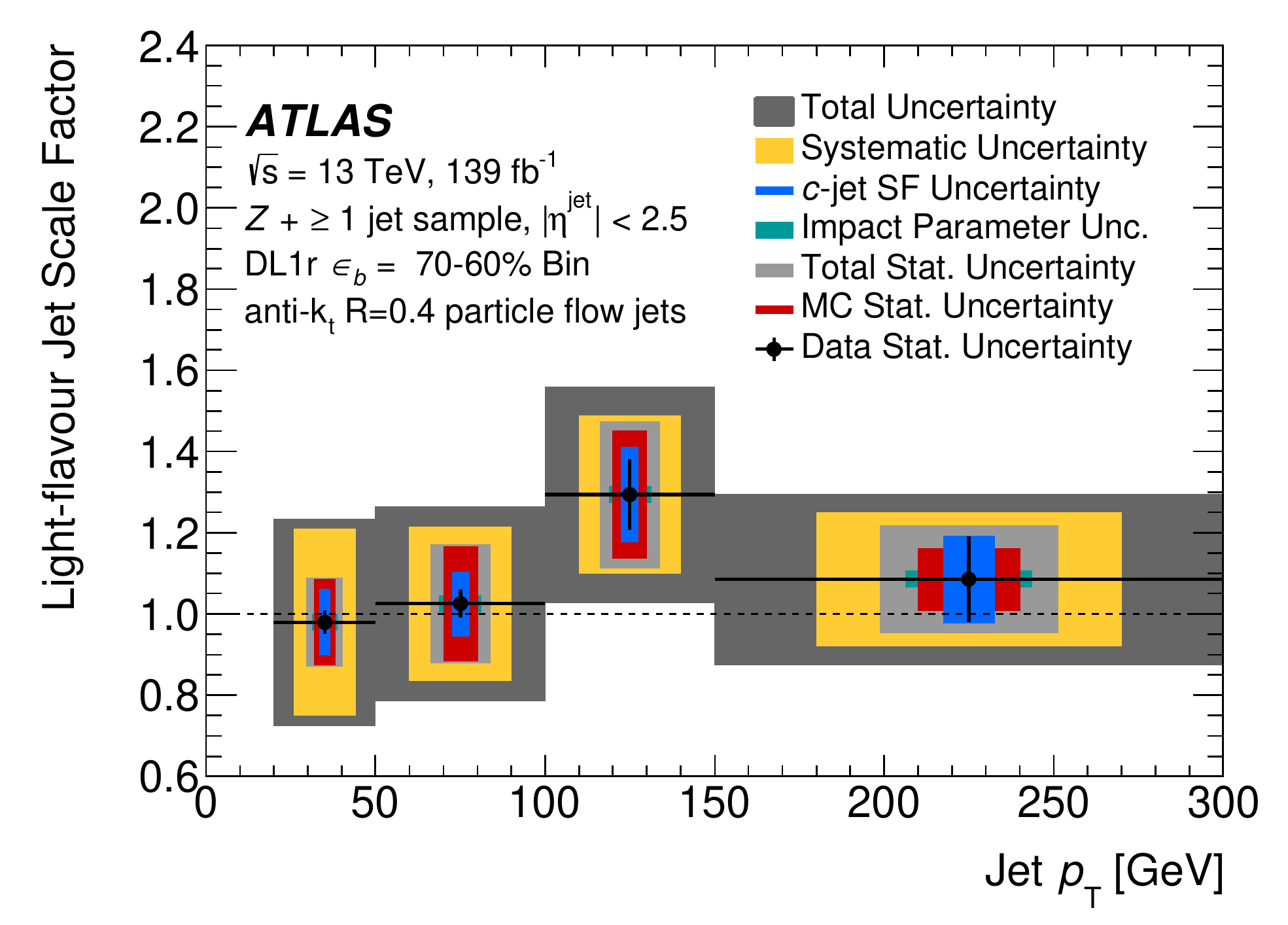}
\includegraphics[width=0.49\textwidth]{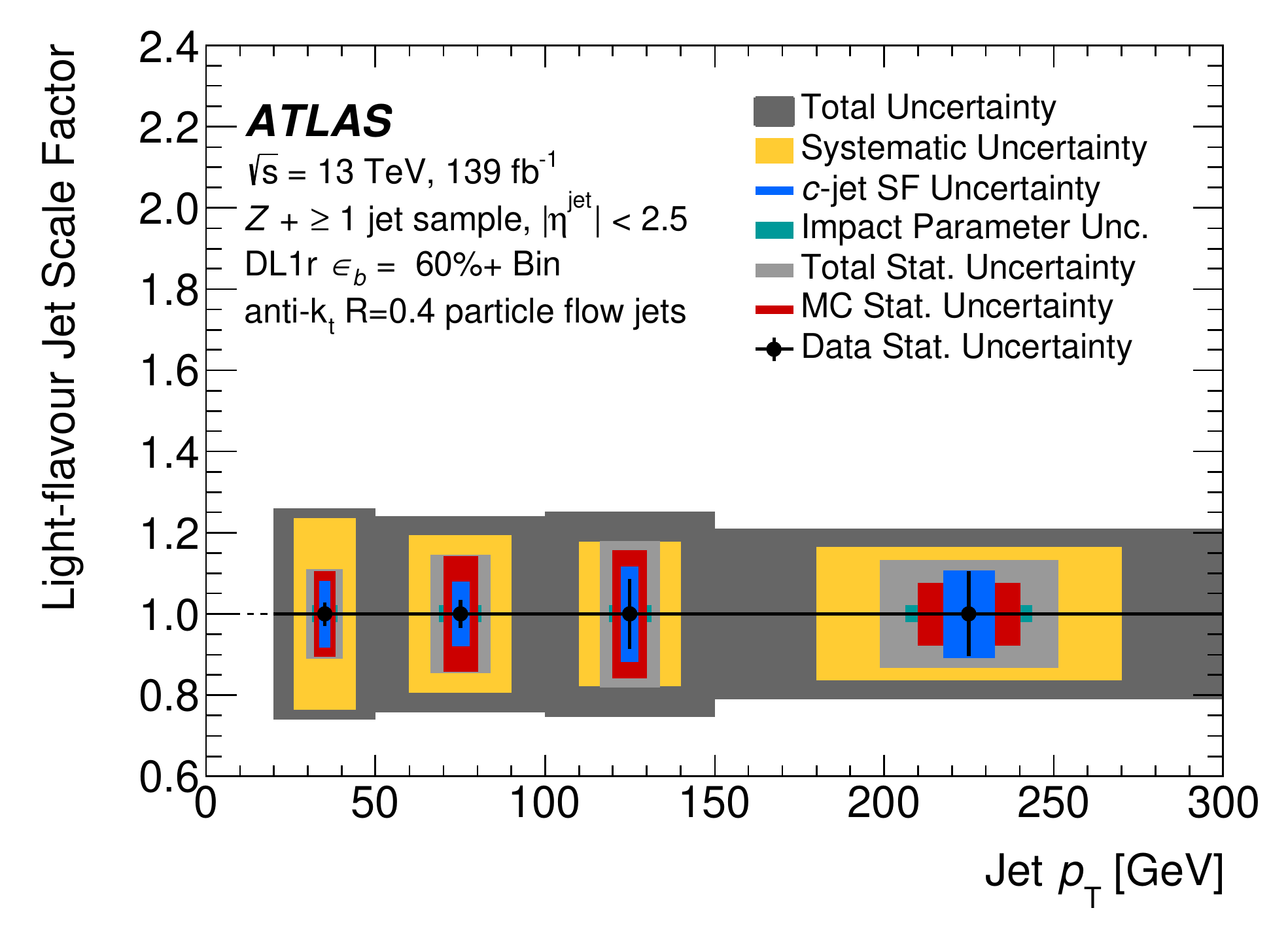} \\
\caption{SFs to calibrate the $\varepsilon^{\text{light}}$ of the pseudo-continuous OPs. The size of the uncertainty bands in the direction of the jet \pT axis is arbitrary and corresponds to the choice of the calibration intervals in \pT.}
\label{fig:DL1rPFlowRun2Results_PC}
\end{figure}

\begin{figure}[h!]
\centering
\includegraphics[width=0.49\textwidth]{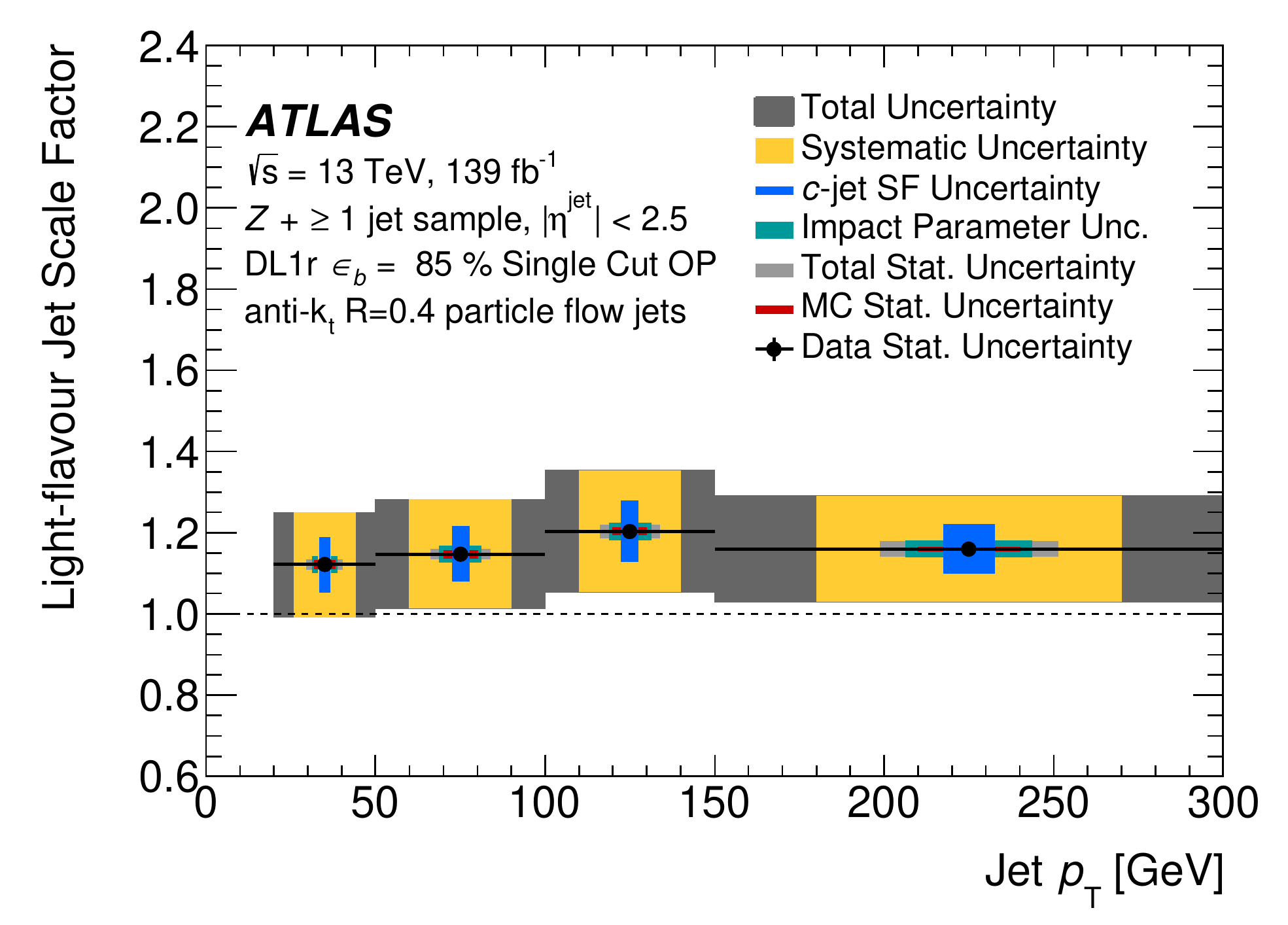}
\includegraphics[width=0.49\textwidth]{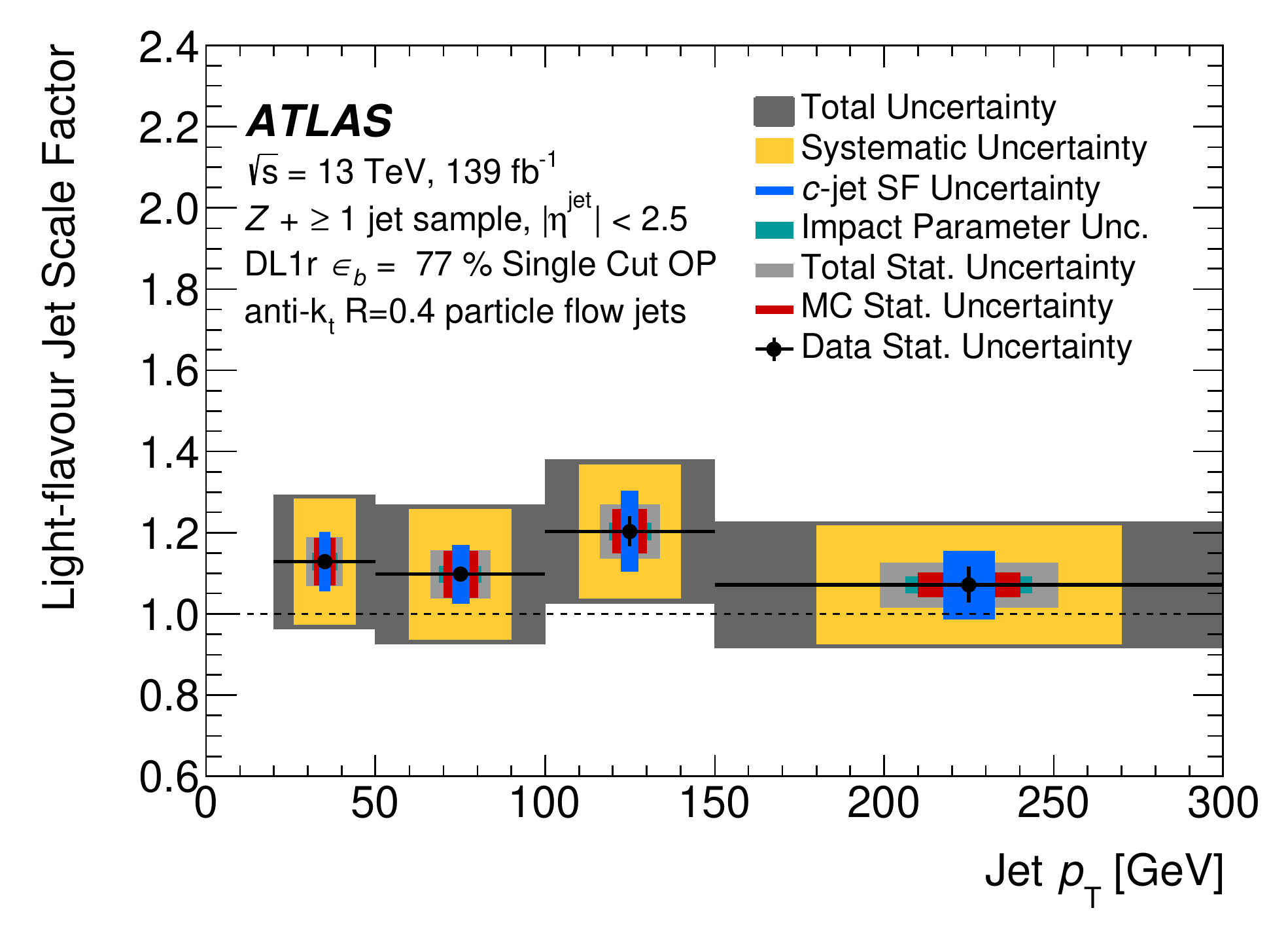} \\
\includegraphics[width=0.49\textwidth]{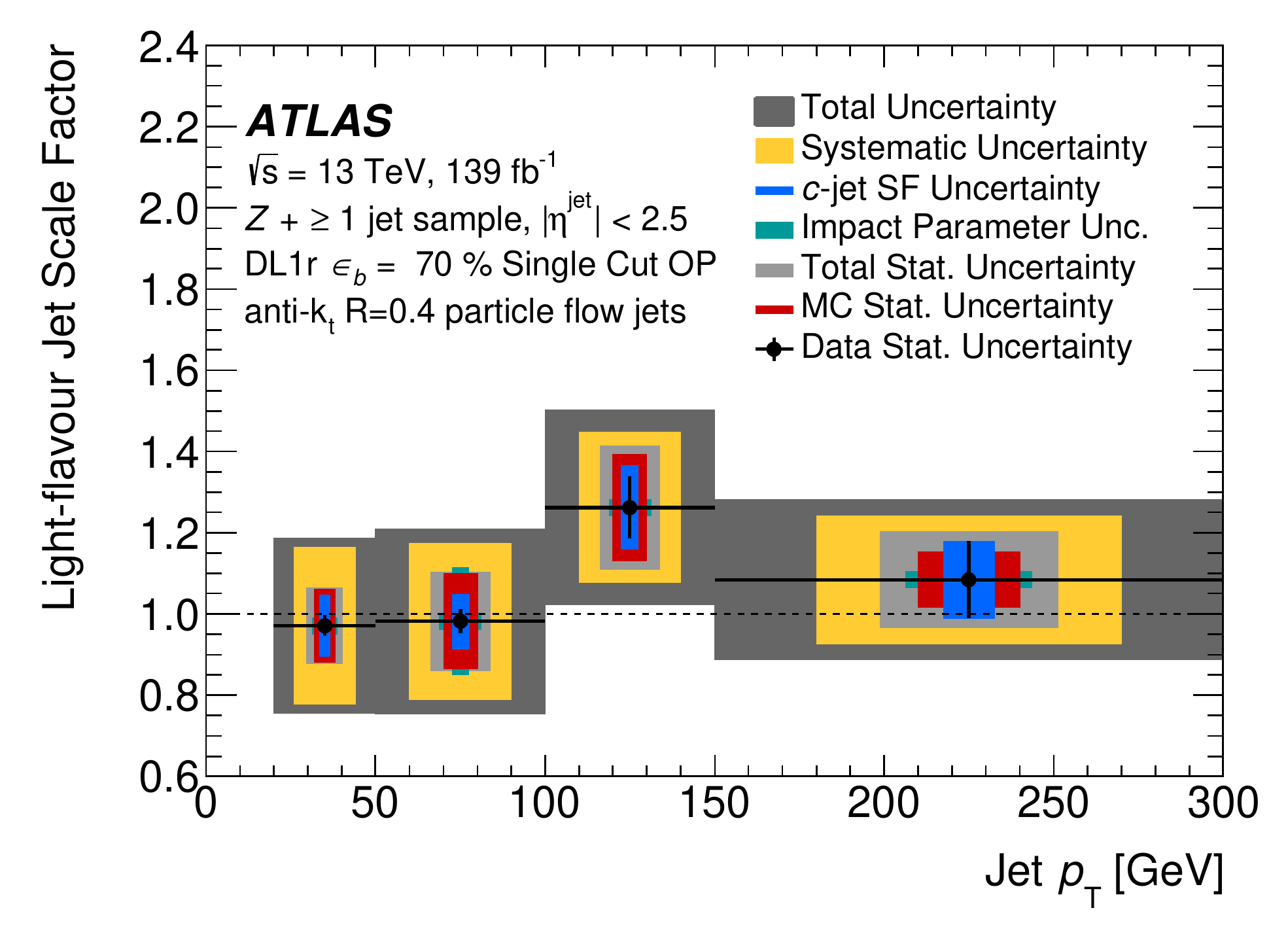}
\includegraphics[width=0.49\textwidth]{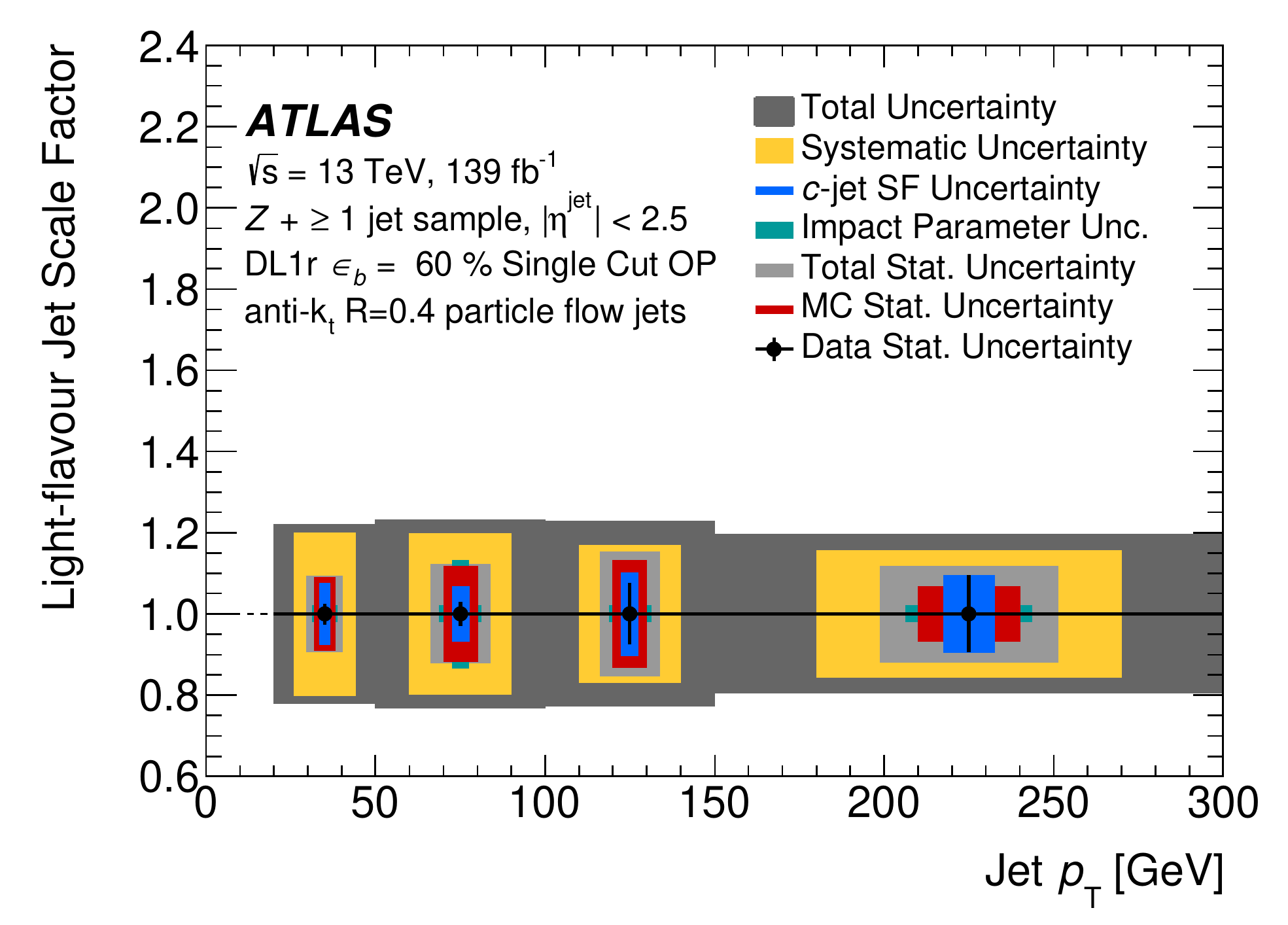}
\caption{SFs to calibrate the $\varepsilon^{\text{light}}$ of the single-cut OPs. The size of the uncertainty bands in the direction of the jet \pT axis is arbitrary and corresponds to the choice of the calibration intervals in \pT.}
\label{fig:DL1rPFlowRun2Results}
\end{figure}
 
\begin{figure}[h!]
\centering
\includegraphics[width=0.49\textwidth]{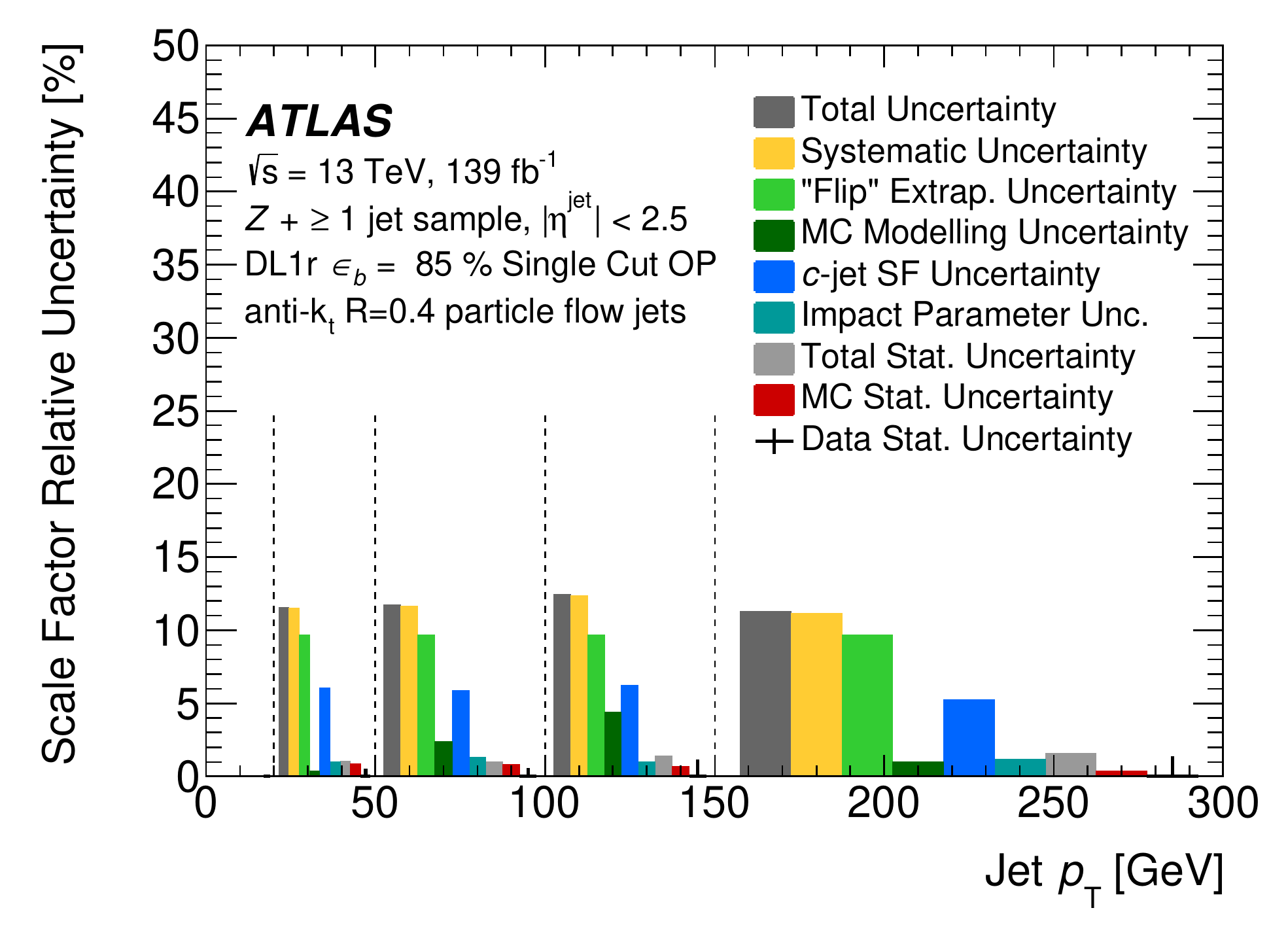}
\includegraphics[width=0.49\textwidth]{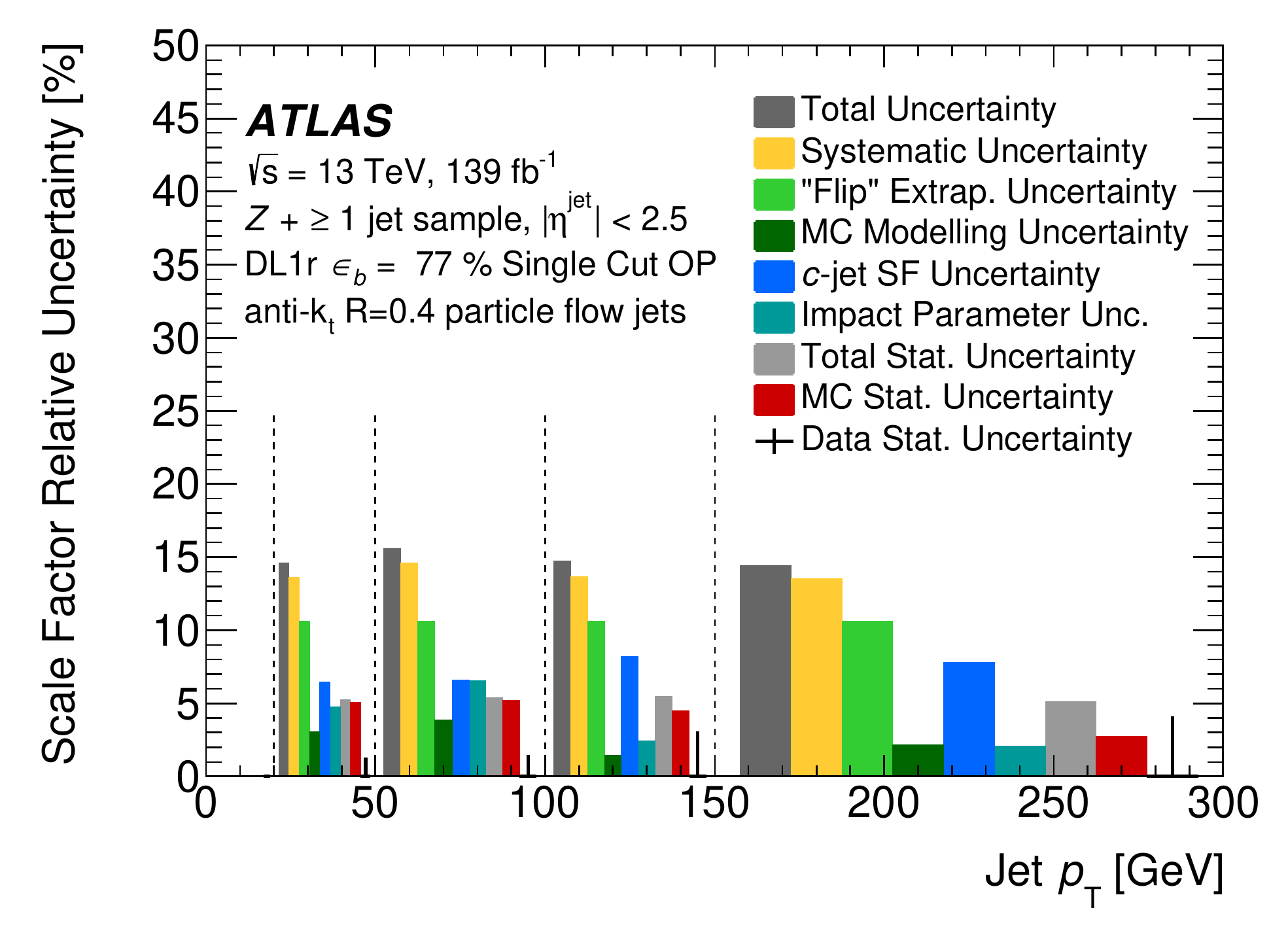} \\
\includegraphics[width=0.49\textwidth]{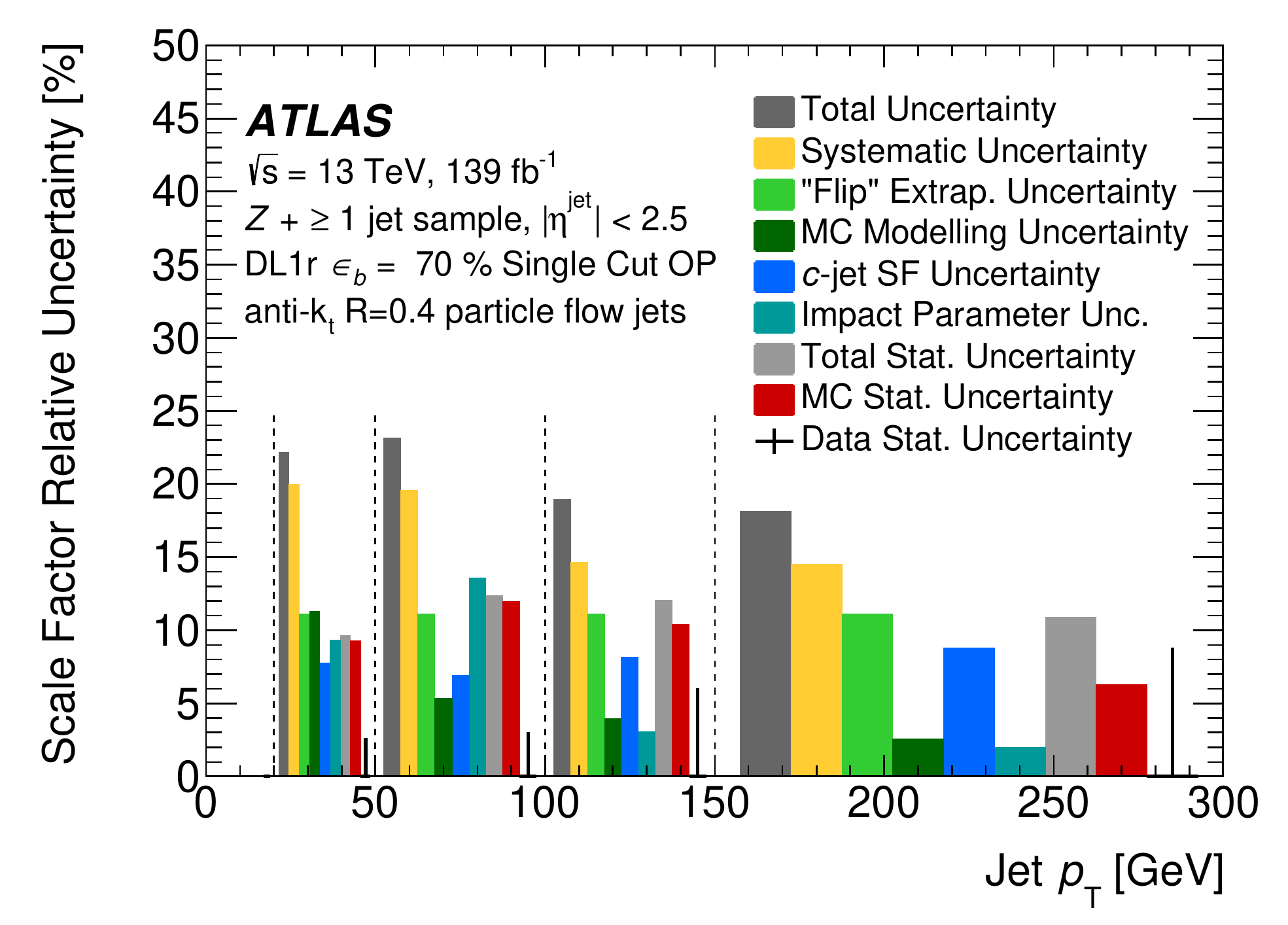}
\includegraphics[width=0.49\textwidth]{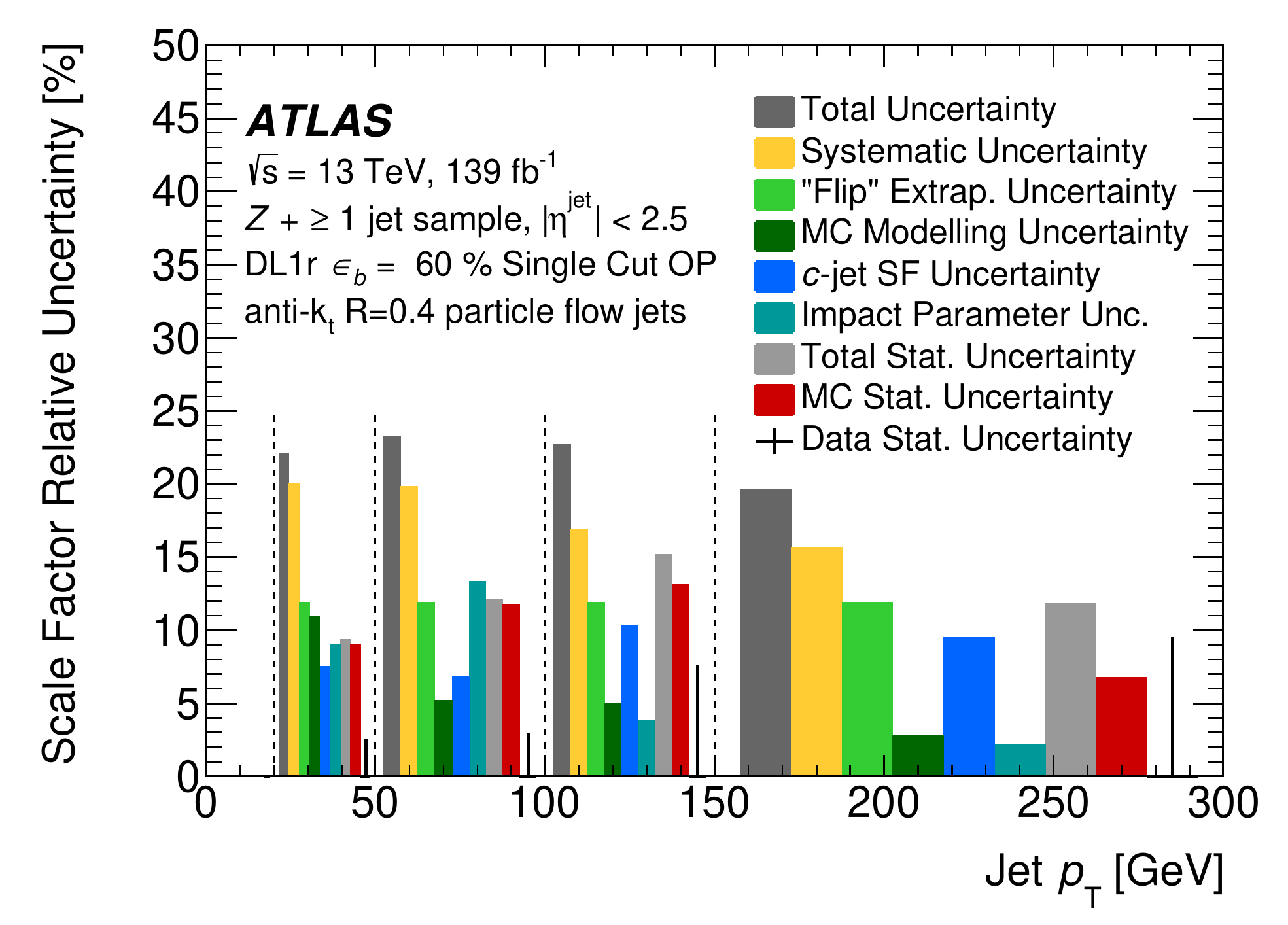}
\caption{\texttt{DL1r} mistagging efficiency calibration uncertainties. The breakdown of the different uncertainties is shown for the different OPs.}
\label{fig:DL1rPFlowRun2Uncertainties}
\end{figure}
\begin{table}[h!]
\caption{Uncertainties on the mistagging efficiency scale factors for the \texttt{DL1r} tagger at the 85\% efficiency OP presented as a function of the jet \pT.}
\label{tab:CaloDL1r_85}

 
{\footnotesize
\begin{center}
\begin{tabular}{l  c  c  c  c  }
\toprule
\multicolumn{5}{c}{Relative uncertainties [\%], 85\%  $\text{SF}^{\text{light}}$ single-cut OP } \\ \hline
$p_{\mathrm{T}}$ bin [\,GeV] & 20 $ < p_{\mathrm{T}} < $ 50 & 50 $ < p_{\mathrm{T}} < $ 100 & 100 $ < p_{\mathrm{T}} < $ 150 & 150 $ < p_{\mathrm{T}} < $ 300\\
\midrule
$c$-jet SF & 6.1 & 5.9 & 6.2 & 5.3\\
`Flip' extrapolation  & 9.7 & 9.7 & 9.7 & 9.7\\
MC modelling  & ~~0.38 & 2.4 & 4.4 & 1~~~ \\
Impact parameter ($d_0, z_0$)  & 1~~~ & 1.3 & 1~~~ & 1.2\\
\midrule
Total systematic uncertainty & 11~~~~~ & 12~~~~~ & 12~~~~~ & 11~~~~~ \\
\midrule
Data statistical & ~~0.52 & ~~0.56 & 1.1 & 1.4\\
MC statistical & ~~0.88 & ~~0.84 & ~~0.72 & 0.4\\
\midrule
Total statistical uncertainty & 1.1 & 1~~~ & 1.4 & 1.6\\
\midrule
Total uncertainty & 12~~~~~ & 12~~~~~ & 12~~~~~ & 11~~~~~ \\
\bottomrule
\end{tabular}
\end{center}
}


\end{table}
 
\begin{table}[h!]
\caption{Uncertainties on the mistagging efficiency scale factors for the \texttt{DL1r} tagger at the 77\% efficiency OP presented as a function of the jet \pT.}

{\footnotesize
\begin{center}
\begin{tabular}{l  c  c  c  c  }
\toprule
\multicolumn{5}{c}{Relative uncertainties [\%], 77\%  $\text{SF}^{\text{light}}$ single-cut OP } \\ \hline
$p_{\mathrm{T}}$ bin [\,GeV] & 20 $ < p_{\mathrm{T}} < $ 50 & 50 $ < p_{\mathrm{T}} < $ 100 & 100 $ < p_{\mathrm{T}} < $ 150 & 150 $ < p_{\mathrm{T}} < $ 300\\
\midrule
$c$-jet SF & 6.4 & 6.6 & 8.2 & 7.8\\
`Flip' extrapolation & 11~~~~~ & 11~~~~~ & 11~~~~~ & 11~~~~~ \\
MC modelling & 3.1 & 3.9 & 1.4 & 2.2\\
Impact parameter ($d_0, z_0$) & 4.8 & 6.5 & 2.4 & 2.1\\
\midrule
Total systematic uncertainty & 14~~~~~ & 15~~~~~ & 14~~~~~ & 14~~~~~ \\
\midrule
Data statistical & 1.3 & 1.5 & 3.1 & 4.1\\
MC statistical & 5.1 & 5.2 & 4.5 & 2.7\\
\midrule
Total statistical uncertainty & 5.2 & 5.4 & 5.5 & 5.1\\
\midrule
Total uncertainty & 15~~~~~ & 16~~~~~ & 15~~~~~ & 14~~~~~ \\
\bottomrule
\end{tabular}
\end{center}
}


\end{table}
 
\begin{table}[h!]
\caption{Uncertainties on the mistagging efficiency scale factors for the \texttt{DL1r} tagger at the 70\% efficiency OP presented as a function of the jet \pT.}

 
{\footnotesize
\begin{center}
\begin{tabular}{l  c  c  c  c  }
\toprule
\multicolumn{5}{c}{Relative uncertainties [\%], 70\%  $\text{SF}^{\text{light}}$ single-cut OP } \\ \hline
$p_{\mathrm{T}}$ bin [\,GeV] & 20 $ < p_{\mathrm{T}} < $ 50 & 50 $ < p_{\mathrm{T}} < $ 100 & 100 $ < p_{\mathrm{T}} < $ 150 & 150 $ < p_{\mathrm{T}} < $ 300\\
\midrule
$c$-jet SF & 7.8 & 6.9 & 8.2 & 8.8\\
`Flip' extrapolation & 11~~~~~ & 11~~~~~ & 11~~~~~ & 11~~~~~ \\
MC modelling & 11~~~~~ & 5.3 & 4~~~ & 2.6\\
Impact parameter ($d_0, z_0$) & 9.3 & 14~~~~~ & 3~~~ & 2~~~ \\
\midrule
Total systematic uncertainty & 20~~~~~ & 20~~~~~ & 15~~~~~ & 15~~~~~ \\
\midrule
Data statistical & 2.6 & 3~~~ & 6~~~ & 8.8\\
MC statistical & 9.3 & 12~~~~~ & 10~~~~~ & 6.3\\
\midrule
Total statistical uncertainty & 9.7 & 12~~~~~ & 12~~~~~ & 11~~~~~ \\
\midrule
Total uncertainty & 22~~~~~ & 23~~~~~ & 19~~~~~ & 18~~~~~ \\
\bottomrule
\end{tabular}
\end{center}
}


\end{table}
 
\begin{table}[h!]
\caption{Uncertainties on the mistagging efficiency scale factors for the \texttt{DL1r} tagger at the 60\% efficiency OP presented as a function of the jet \pT.}
\label{tab:CaloDL1r_60}

{\footnotesize
\begin{center}
\begin{tabular}{l  c  c  c  c  }
\toprule
\multicolumn{5}{c}{Relative uncertainties [\%], 60\%  $\text{SF}^{\text{light}}$ single-cut OP } \\ \hline
$p_{\mathrm{T}}$ bin [\,GeV] & 20 $ < p_{\mathrm{T}} < $ 50 & 50 $ < p_{\mathrm{T}} < $ 100 & 100 $ < p_{\mathrm{T}} < $ 150 & 150 $ < p_{\mathrm{T}} < $ 300\\
\midrule
$c$-jet SF & 7.6 & 6.8 & 10~~~~~ & 9.5\\
`Flip' extrapolation & 12~~~~~ & 12~~~~~ & 12~~~~~ & 12~~~~~ \\
MC modelling & 11~~~~~ & 5.2 & 5~~~ & 2.8\\
Impact parameter ($d_0, z_0$) & 9.1 & 13~~~~~ & 3.8 & 2.2\\
\midrule
Total systematic uncertainty & 20~~~~~ & 20~~~~~ & 17~~~~~ & 16~~~~~ \\
\midrule
Data statistical & 2.6 & 3~~~ & 7.6 & 9.5\\
MC statistical & 9~~~ & 12~~~~~ & 13~~~~~ & 6.8\\
\midrule
Total statistical uncertainty  & 9.4 & 12~~~~~ & 15~~~~~ & 12~~~~~ \\
\midrule
Total uncertainty & 22~~~~~ & 23~~~~~ & 23~~~~~ & 20~~~~~ \\
\bottomrule
\end{tabular}
\end{center}
}


\end{table}
 
In the previous calibration with the Negative Tag method using dijet events in 2015--2016 data, the total uncertainty in SF$^{\text{light}}$ ranged from 14\% to 76\% for jets with \pT between 20 and 300~GeV. Dominant uncertainties arose from the modelling of the track impact parameter resolution and the uncertainty in the fraction of heavy-flavour jets which was estimated using simulation. The reduced uncertainty in the present calibration stems from both improved modelling of the track impact parameters in simulation and the $m_{\text{SV}}$ fit, which allows the extraction of SF$^{\text{light}}$ while simultaneously correcting $\varepsilon^{b}$ and the jet-flavour fractions using data. Previously, $\varepsilon^{b}$ and the heavy-flavour fractions were estimated from simulation, resulting in large modelling uncertainties, which were a limiting factor in the precision of the result.


\section{Conclusions}
\label{sec:conclusions}

The light-flavour jet mistagging efficiency $\varepsilon^{\text{light}}$ of the \texttt{DL1r} $b$-tagging algorithm has been measured with a $139$~$\ifb$ sample of $\rts=13~\TeV$ $pp$ collision events recorded during 2015--2018 by the ATLAS detector at the LHC.
The measurement is based on an improved method applied to a sample of $Z$+jets events.
The Negative Tag method, based on the application of an alternative $b$-tagging algorithm, designed to facilitate the measurement of the light-flavour jet mistagging efficiency, is used.
Data-to-simulation scale factors for correcting $\varepsilon^{\text{light}}$ in simulation are measured in four different jet transverse momentum intervals, ranging from 20 to 300~\GeV, for four separate quantiles of the $b$-tagging discriminant.
The scale factors typically exceed unity by around $10\%$--$20\%$, with total uncertainties ranging from $11\%$ to $23\%$, and do not exhibit any strong dependence on jet transverse momentum.
These calibration uncertainties are considerably lower than the previous $14\%$ to $76\%$ uncertainties from the Negative Tag method using 2015--2016 data.


\FloatBarrier

\section*{Acknowledgements}


We thank CERN for the very successful operation of the LHC, as well as the
support staff from our institutions without whom ATLAS could not be
operated efficiently.
 
We acknowledge the support of
ANPCyT, Argentina;
YerPhI, Armenia;
ARC, Australia;
BMWFW and FWF, Austria;
ANAS, Azerbaijan;
CNPq and FAPESP, Brazil;
NSERC, NRC and CFI, Canada;
CERN;
ANID, Chile;
CAS, MOST and NSFC, China;
Minciencias, Colombia;
MEYS CR, Czech Republic;
DNRF and DNSRC, Denmark;
IN2P3-CNRS and CEA-DRF/IRFU, France;
SRNSFG, Georgia;
BMBF, HGF and MPG, Germany;
GSRI, Greece;
RGC and Hong Kong SAR, China;
ISF and Benoziyo Center, Israel;
INFN, Italy;
MEXT and JSPS, Japan;
CNRST, Morocco;
NWO, Netherlands;
RCN, Norway;
MEiN, Poland;
FCT, Portugal;
MNE/IFA, Romania;
MESTD, Serbia;
MSSR, Slovakia;
ARRS and MIZ\v{S}, Slovenia;
DSI/NRF, South Africa;
MICINN, Spain;
SRC and Wallenberg Foundation, Sweden;
SERI, SNSF and Cantons of Bern and Geneva, Switzerland;
MOST, Taiwan;
TENMAK, T\"urkiye;
STFC, United Kingdom;
DOE and NSF, United States of America.
In addition, individual groups and members have received support from
BCKDF, CANARIE, Compute Canada and CRC, Canada;
PRIMUS 21/SCI/017 and UNCE SCI/013, Czech Republic;
COST, ERC, ERDF, Horizon 2020 and Marie Sk{\l}odowska-Curie Actions, European Union;
Investissements d'Avenir Labex, Investissements d'Avenir Idex and ANR, France;
DFG and AvH Foundation, Germany;
Herakleitos, Thales and Aristeia programmes co-financed by EU-ESF and the Greek NSRF, Greece;
BSF-NSF and MINERVA, Israel;
Norwegian Financial Mechanism 2014-2021, Norway;
NCN and NAWA, Poland;
La Caixa Banking Foundation, CERCA Programme Generalitat de Catalunya and PROMETEO and GenT Programmes Generalitat Valenciana, Spain;
G\"{o}ran Gustafssons Stiftelse, Sweden;
The Royal Society and Leverhulme Trust, United Kingdom.
 
The crucial computing support from all WLCG partners is acknowledged gratefully, in particular from CERN, the ATLAS Tier-1 facilities at TRIUMF (Canada), NDGF (Denmark, Norway, Sweden), CC-IN2P3 (France), KIT/GridKA (Germany), INFN-CNAF (Italy), NL-T1 (Netherlands), PIC (Spain), ASGC (Taiwan), RAL (UK) and BNL (USA), the Tier-2 facilities worldwide and large non-WLCG resource providers. Major contributors of computing resources are listed in Ref.~\cite{ATL-SOFT-PUB-2021-003}.


\clearpage
\appendix

\printbibliography
 
\clearpage
\input{atlas_authlist}

\end{document}

%% file: atlas_authlist.tex
 
\begin{flushleft}
\hypersetup{urlcolor=black}
{\Large The ATLAS Collaboration}

\bigskip

\AtlasOrcid[0000-0002-6665-4934]{G.~Aad}$^\textrm{\scriptsize 102}$,
\AtlasOrcid[0000-0002-5888-2734]{B.~Abbott}$^\textrm{\scriptsize 120}$,
\AtlasOrcid[0000-0002-7248-3203]{D.C.~Abbott}$^\textrm{\scriptsize 103}$,
\AtlasOrcid[0000-0002-1002-1652]{K.~Abeling}$^\textrm{\scriptsize 55}$,
\AtlasOrcid[0000-0002-8496-9294]{S.H.~Abidi}$^\textrm{\scriptsize 29}$,
\AtlasOrcid[0000-0002-9987-2292]{A.~Aboulhorma}$^\textrm{\scriptsize 35e}$,
\AtlasOrcid[0000-0001-5329-6640]{H.~Abramowicz}$^\textrm{\scriptsize 151}$,
\AtlasOrcid[0000-0002-1599-2896]{H.~Abreu}$^\textrm{\scriptsize 150}$,
\AtlasOrcid[0000-0003-0403-3697]{Y.~Abulaiti}$^\textrm{\scriptsize 117}$,
\AtlasOrcid[0000-0003-0762-7204]{A.C.~Abusleme~Hoffman}$^\textrm{\scriptsize 137a}$
% ...
\bigskip
\\

% ...
$^{29}$Physics Department, Brookhaven National Laboratory, Upton NY; United States of America.\\
% ...
$^{35}$$^{(a)}$Facult\'e des Sciences Ain Chock, R\'eseau Universitaire de Physique des Hautes Energies - Universit\'e Hassan II, Casablanca;$^{(b)}$Facult\'{e} des Sciences, Universit\'{e} Ibn-Tofail, K\'{e}nitra;$^{(c)}$Facult\'e des Sciences Semlalia, Universit\'e Cadi Ayyad, LPHEA-Marrakech;$^{(d)}$LPMR, Facult\'e des Sciences, Universit\'e Mohamed Premier, Oujda;$^{(e)}$Facult\'e des sciences, Universit\'e Mohammed V, Rabat;$^{(f)}$Institute of Applied Physics, Mohammed VI Polytechnic University, Ben Guerir; Morocco.\\
% ...
$^{55}$II. Physikalisches Institut, Georg-August-Universit\"{a}t G\"ottingen, G\"ottingen; Germany.\\
% ...
$^{102}$CPPM, Aix-Marseille Universit\'e, CNRS/IN2P3, Marseille; France.\\
$^{103}$Department of Physics, University of Massachusetts, Amherst MA; United States of America.\\
% ...
$^{117}$Department of Physics, New York University, New York NY; United States of America.\\
% ...
$^{120}$Homer L. Dodge Department of Physics and Astronomy, University of Oklahoma, Norman OK; United States of America.\\
% ...
$^{137}$$^{(a)}$Departamento de F\'isica, Pontificia Universidad Cat\'olica de Chile, Santiago;$^{(b)}$Millennium Institute for Subatomic physics at high energy frontier (SAPHIR), Santiago;$^{(c)}$Instituto de Investigaci\'on Multidisciplinario en Ciencia y Tecnolog\'ia, y Departamento de F\'isica, Universidad de La Serena;$^{(d)}$Universidad Andres Bello, Department of Physics, Santiago;$^{(e)}$Instituto de Alta Investigaci\'on, Universidad de Tarapac\'a, Arica;$^{(f)}$Departamento de F\'isica, Universidad T\'ecnica Federico Santa Mar\'ia, Valpara\'iso; Chile.\\
% ...
$^{150}$Department of Physics, Technion, Israel Institute of Technology, Haifa; Israel.\\
$^{151}$Raymond and Beverly Sackler School of Physics and Astronomy, Tel Aviv University, Tel Aviv; Israel.\\
% ...
$^{a}$ Also Affiliated with an institute covered by a cooperation agreement with CERN.\\
$^{b}$ Also at An-Najah National University, Nablus; Palestine.\\
$^{c}$ Also at Borough of Manhattan Community College, City University of New York, New York NY; United States of America.\\
$^{d}$ Also at Bruno Kessler Foundation, Trento; Italy.\\
$^{e}$ Also at Center for High Energy Physics, Peking University; China.\\
$^{f}$ Also at Center for Interdisciplinary Research and Innovation (CIRI-AUTH), Thessaloniki ; Greece.\\
% ...

\end{flushleft}
